\documentclass{aa}
\usepackage{amsmath,graphicx,subfigure,natbib,url,txfonts, mathtools,spverbatim}
\bibpunct{(}{)}{;}{a}{}{,}

\newcommand{\ie}
  {{\em i.e.}}
\newcommand{\eg}
  {{\em e.g.}}
\newcommand{\minimum}
  {{\rm{min}}}
\newcommand{\maximum}
  {{\rm{max}}}
\newcommand{\step}
  {\Theta}
\newcommand{\sumband}
  {{\sum_{\band = 1}^{\nband}}}
\newcommand{\prodband}
  {{\prod_{\band = 1}^{\nband}}}
\newcommand{\offset}
  {m}

\newcommand{\diff}
  {{\rm{d}}}
\newcommand{\prob}
  {{\rm{Pr}}}

\newcommand{\data}
  {\{ \est{m}_\band \}}
\newcommand{\noise}
  {\{ \sigma_\band \}}
\newcommand{\nparam} 
  {{N_{\rm{\theta}}}}
\newcommand{\params}
  {\theta_1, \theta_2, \ldots , \theta_\nparam}

\newcommand{\band}
  {b}
\newcommand{\nband}
  {N_{\rm{b}}}

\newcommand{\ntype}
  {{N_{\rm{t}}}}

\newcommand{\est}
  {\hat}

\begin{document}

\title{Photometric brown-dwarf classification. I.} 

\subtitle{A method to identify and accurately classify large samples 
  of brown dwarfs without spectroscopy} 

\author{Skrzypek, N.\inst{\ref{inst1}}, Warren,
  S. J.\inst{\ref{inst1}}, Faherty, J.K.\inst{\ref{inst2}}, Mortlock,
  D. J.\inst{\ref{inst1},\ref{inst3}}, Burgasser,
  A.J.\inst{\ref{inst4}}, Hewett, P.C.\inst{\ref{inst5}}} \authorrunning{Skrzypek, N., et
  al.}

\institute{Astrophysics Group, Imperial College London, Blackett
  Laboratory, Prince Consort Road, London SW7 2AZ, UK \label{inst1}
  \and Department of Terrestrial Magnetism, Carnegie Institution of
  Washington, Washington, DC 20015, USA \label{inst2} \and Department
  of Mathematics, Imperial College London, London SW7 2AZ,
  UK \label{inst3} \and Department of Physics, University of
  California, San Diego, CA 92093, USA \label{inst4} \and Institute of
Astronomy, Madingley Road, Cambridge CB3 0HA, UK \label{inst5}}

\date{Received 2014 July 9 / Accepted 2014 November 27}

\abstract{}{We present a method, named {\em photo-type}, to identify
  and accurately classify L and T dwarfs onto the standard spectral
  classification system using photometry alone. This enables the
  creation of large and deep homogeneous samples of these objects
  efficiently, without the need for spectroscopy.}{We created a
  catalogue of point sources with photometry in 8 bands, ranging from
  0.75 to 4.6 $\mu$m, selected from an area of 3344 $\deg^2$, by
  combining SDSS, UKIDSS LAS, and WISE data. Sources with
  $13.0<J<17.5$, and $Y-J>0.8$, were then classified by comparison
  against template colours of quasars, stars, and brown dwarfs. The L
  and T templates, spectral types L0 to T8, were created by
  identifying previously known sources with spectroscopic
  classifications, and fitting polynomial relations between colour and
  spectral type.}{Of the 192 known L and T dwarfs with reliable
  photometry in the surveyed area and magnitude range, 189 are
  recovered by our selection and classification method. We have
  quantified the accuracy of the classification method both
  externally, with spectroscopy, and internally, by creating synthetic
  catalogues and accounting for the uncertainties. We find that,
  brighter than $J=17.5$, $p$$h$$o$$t$$o$-$t$$y$$p$$e$ classifications are
  accurate to one spectral sub-type, and are therefore competitive
  with spectroscopic classifications. The resultant catalogue of 1157
  L and T dwarfs will be presented in a companion paper.}{}

\keywords{Stars: low-mass -- Techniques:
  photometric -- Methods: data analysis -- Stars: individual:  
  SDSS~J1030+0213, 2MASS~J1542-0045, ULAS~J2304+1301} 

\maketitle

\section{Introduction}

The first brown dwarfs were discovered by \citet{nakajima95} and
\citet{rebolo95}, having been theorised earlier by
\citet{kumar1,kumar2} and \citet{hayashi}.  Exploration of the brown
dwarf population has proceeded rapidly over the subsequent two
decades, enabled by new surveys in the optical, the near-infrared, and
the mid-infrared. This has resulted in the creation of three new,
successively cooler, spectral classes beyond M: the L
\citep{kirkpatrick99,martin1999}; T
\citep{geballe02,burgasser02,burgasser06b}; and Y dwarfs
\citep{cushing11}. The temperature sequence has now been mapped all
the way down to effective temperatures of $\sim250$ K
\citep{luhman14}. This almost closes the gap to the effective
temperature of gas giants in our Solar System ($\sim100$ K). The
current paper focuses on L and T dwarfs.

One of the fundamental observables that characterises the LTY
population is the luminosity function (e.g. \citealt{Cruz07, Reyle10})
i.e. the dependence of space density on absolute magnitude (or,
equivalently, spectral type). The luminosity function can be used to
learn about the sub-stellar initial mass function and birth
rate. The study of the luminosity function requires a homogeneous
sample with well-defined selection criteria. 

DwarfArchives.org provides a compilation of L and T dwarfs published
in the literature. This collection is heterogeneous, having been culled
from several surveys with different characteristics, including, in
particular, the Sloan Digital Sky Survey (SDSS; \citealt{york00}) in
the optical, the Deep Near-Infrared Southern Sky Survey (DENIS;
\citealt{epchtein97}), the Two Micron All-Sky Survey (2MASS;
\citealt{2MASS}) and the UKIRT Infrared Deep Sky Survey (UKIDSS;
\citealt{UKIDSS}) in the near-infrared, and WISE in the
mid-infrared. DwarfArchives.org now contains over 1000 L and T dwarfs
but the constituent samples are themselves heterogeneous and so not
well suited for statistical analysis.

A good example of the state of the art in obtaining a statistical
sample of brown dwarfs is the study of the sub-stellar birth rate by
\citet{DayJones13}.  They classified 63 new L and T dwarfs brighter
than $J=18.1$, using X-shooter spectroscopy on the Very Large
Telescope (VLT), after selecting them through colour cuts.  The sample
targeted a limited section of the L and T sequence and required
substantial follow-up resources: at $J\simeq17.5$ a near-infrared
spectrum good enough for classification to one spectral sub-type
requires about 30 min on an 8 m class telescope.

In this paper we consider how to define a homogeneous, complete sample
of field L and T dwarfs, with accurate classifications, over the full
L and T spectral range, from L0 to T8, that reaches similar depth and
that is more than an order of magnitude larger. Such a sample is
useful to reduce the uncertainty in the measurement of the luminosity
function and also allows a variety of studies, such as measuring the
brown-dwarf scale height \citep{ryan05,juric08}, the frequency of
binarity \citep{burgasser06a,burgasser07a,luhman12} and, if proper
motions can be measured, kinematic studies
\citep{faherty09,faherty12,schmidt10,smith14}. A large sample can also
be used to identify rare, unusual objects (e.g. \citealt{burgasser03sub},
\citealt{folkes07}, \citealt{looper08}).

All previous searches for L and T dwarfs have required spectroscopy
for accurate classification. This paper describes an alternative
search and classification method that uses only existing photometric
survey data. We call the method {\em photo-type}, by analogy with {\em
  photo-z}, the measurement of galaxy redshifts from photometric data
alone. \S\ref{sec:technique} sets out the details of the technique. The
method of classification, explained in \S\ref{subsec:classification},
requires comparison of the multiband photometry of the object against
a set of template colours, to find the best fit. The creation of the
template set is explained in \S\ref{subsec:templates}. In
\S\ref{subsec:bias} we discuss possible sources of bias in the
template colours. In \S\ref{subsec:binaries} we additionally consider
the possibility of identifying unresolved binaries by their unusual
colours. We then apply the new technique in \S\ref{sec:application}.
In \S\ref{subsec:photom} we describe the creation of a catalogue of
multi wavelength photometry of point sources with matched photometry
in SDSS, UKIDSS, and WISE. The results of a search of this catalogue
for L and T dwrafs are provided in Paper II. In
\S\ref{subsec:accuracy} we quantify the accuracy of the classification
method. \S\ref{sec:cook} provides a ``cookbook'' for {\em photo-type},
explaining how to measure the spectral type, and its accuracy, for a
source with photometry in all or a subset of the 8 bands used here. We
summarise in \S\ref{sec:summary}. In Paper II we present the catalogue
of 1157 L and T dwarfs classified by {\em photo-type}, and quantify
the completeness of the sample.

The photometric bands used in this study are the $i$ and $z$ bands in
SDSS, the $Y$, $J$, $H$, $K$ bands in UKIDSS, and the $W1$, $W2$ bands
in WISE. {\em All the magnitudes and colours quoted in this paper are
  Vega based.} The $YJHKW1W2$ survey data are calibrated to Vega,
while SDSS is calibrated on the AB system; We have applied the offsets
tabulated in \citet{hewett06} to convert the SDSS $iz$ AB magnitudes
to Vega.

\section{The technique}
\label{sec:technique}

\subsection{Classification by $\chi^2$}
\label{subsec:classification}


The overall aim here is to develop a method to classify every
point-source in a multiband photometric catalogue as a star, white
dwarf, brown dwarf or quasar, based only on its observed photometry.
This could be achieved most rigorously by using Bayesian model
comparison \citep{mortlock12}.  A formalism for doing so is detailed
in the Appendix, which also includes the series of approximations and
assumptions which lead from the full Bayesian result to the much
simpler $\chi^2$-based classification we actually use.  While some of
the approximations below are clearly unrealistic, that is unimportant
per se -- all that matters in this context is whether the final
classifications remain (largely) unchanged.  The effect of our
approximations are examined in detail at the end of
\S\ref{subsec:templates}, but justified more qualitatively here.

The first approximation is that we work with magnitudes and assume the
photometric errors are Gaussian. This is justified at high
signal-to-noise ratios (S/N). At low S/N, the correct approach is to
work with fluxes (e.g., \citealt{mortlock12}) where, except in the
low-photon regime (e.g., X-ray astronomy), the errors are Gaussian. A
few T dwarfs in our sample have low S/N in the SDSS $i$ and $z$ bands;
for these we use |{\em asinh}| magnitudes and errors \citep{lupton99}
as the resultant $\chi^2$ value is close to that which would have been
calculated from the fluxes.

The second simplification is to adopt equal prior probabilities for
each of the different (sub-)types. In the limited region of
colour-magnitude space that we search for L and T dwarfs,
$13.0<J<17.5$, $Y-J>0.8$, the main contaminating population is
reddened quasars (see \citealt{hewett06}). At these bright magnitudes
this population is easily distinguished, i.e. the priors are not very
important.  Similarly, provided the priors vary relatively slowly
along the MLT spectral sequence and the likelihoods are sharply
peaked, the prior probabilities for all the sub-types can be assumed
to be equal without affecting the final classifications.

A third, related, approximation is to adopt broad, uniform priors for
the magnitude distributions (i.e., number counts) of each type. This
again is obviously unrealistic, as the number of L and T dwarfs per magnitude is
expected to be $N(m) \propto 10^{3m/5}$, as they are an approximately
uniform population in locally Euclidean space.  But the form of this
prior has minimal impact on the classifications because we are
operating in the high S/N regime in which the data constrain the
overall flux-level of a source to a narrow range.

Applying the above assumptions and approximations to the full Bayesian
formalism, described in the Appendix, results in the following
classification scheme:

\begin{enumerate}

\item
The first requirement is photometric data on the source to be
classified: the measured magnitudes, $\{\hat{m}_b\}$, and
uncertainties, $\{\sigma_b\}$, in each of $\nband$ bands (with $b \in
\{1, 2, \ldots, \nband\}$)\footnote{The braces $\{\}$ are used to denote a list of values, so that, e.g.,
 $\{m_b\} = \{m_1, m_2, \ldots, m_{N_{\rm b}}\}$ is the list of true
 magnitudes in each of the $N_{\rm b}$ bands.  This is not a set, as 
 it is ordered, but neither is it a vector as it is not a geometrical
 object.}.
  For most sources considered here $\nband = 8$ and the filters are
  $izYJHKW1W2$.

\item
The other required input is a set of $N_t$ source types, each specified
by a set of template colours, $\{c_{b,t}\}$, which give the magnitude
difference between band $b$ and some reference band $B$ for objects of
type $t$ (with $t \in \{1, 2, \ldots, N_t\}$).  Hence $c_{B,t} = 0$ by
construction, meaning that the magnitude in this reference band,
$m_B$, is the natural quantity to specify the overall source
brightness.  Here we use $J$ as the reference band as our
conservative magnitude cuts in this band ensure that $m_B = m_J$ is
well constrained.

\item
The first processing step is, for each of the $N_t$ types, to 
calculate the inverse variance weighted estimate of the 
reference magnitude,
\begin{equation}
\hat{m}_{B,t}
=\frac{
  \displaystyle\sum\limits_{b=1}^{N_b} 
    \frac{\hat{m}_b-c_{b,t}} {\sigma_b^2}
  }
  {
    \displaystyle\sum\limits_{b=1}^{N_b}\frac{1}{\sigma_b^2}
  }
  ,
\label{weighted_mean}
\end{equation}
which, in general, is different for each template. 
\item
Next, the above value for $\hat{m}_{B,t}$ is used to calculate 
the minimum $\chi^2$ value for type $t$,
\begin{equation}
\chi^2(\{\hat{m}_b\},\{\sigma_b\},\hat{m}_{B,t},t)
=
\sum_{b=1}^{N_b}\Bigg(\frac{\hat{m}_b-\hat{m}_{B,t}-c_{b,t}}{\sigma_b}\Bigg)^2.
\label{chi2_eqn}
\end{equation}
It is important that $\chi^2$ is calculated by comparing measured
and predicted magnitudes, as opposed to colours.  The reasons for  
this are given in the Appendix.
 
\item
Finally, the source is classified as being of the type $t$ which
results in the smallest value of
$\chi^2(\{\hat{m}_b\},\{\sigma_b\},\hat{m}_{B,t},t)$.  The conditions
under which this corresponds to the most probable type are detailed in
the Appendix; a more empirical demonstration that this results in a
reliable classification is given in \S4.

\end{enumerate}



\subsection{Templates}
\label{subsec:templates}

\begin{table}\footnotesize
\centering
\begin{tabular}{l l l l l l l l }
\hline\hline
{SpT} & {$i-z$} & {$z-Y$} & {$Y-J$} & {$J-H$} & {$H-K$} & {$K-W1$} & {$W1-W2$}\\\hline 
M5 & 0.91 & 0.47 & 0.55 & 0.45 & 0.32 & 0.11 & 0.17\\
M6 & 1.45 & 0.60 & 0.67 & 0.53 & 0.39 & 0.22 & 0.21\\
M7 & 1.77 & 0.70 & 0.78 & 0.56 & 0.44 & 0.25 & 0.24\\
M8 & 1.93 & 0.77 & 0.87 & 0.58 & 0.47 & 0.26 & 0.26\\
M9 & 1.99 & 0.82 & 0.96 & 0.60 & 0.51 & 0.27 & 0.27\\
L0 & 2.01 & 0.86 & 1.04 & 0.63 & 0.54 & 0.29 & 0.27\\
L1 & 2.02 & 0.88 & 1.11 & 0.67 & 0.58 & 0.33 & 0.28\\
L2 & 2.04 & 0.90 & 1.18 & 0.73 & 0.63 & 0.40 & 0.28\\
L3 & 2.10 & 0.92 & 1.23 & 0.79 & 0.67 & 0.48 & 0.29\\
L4 & 2.20 & 0.94 & 1.27 & 0.86 & 0.71 & 0.56 & 0.30\\
L5 & 2.33 & 0.97 & 1.31 & 0.91 & 0.74 & 0.65 & 0.32\\
L6 & 2.51 & 1.00 & 1.33 & 0.96 & 0.75 & 0.72 & 0.36\\
L7 & 2.71 & 1.04 & 1.35 & 0.97 & 0.75 & 0.77 & 0.41\\
L8 & 2.93 & 1.09 & 1.21 & 0.96 & 0.71 & 0.79 & 0.48\\
L9 & 3.15 & 1.16 & 1.20 & 0.90 & 0.65 & 0.79 & 0.57\\
T0 & 3.36 & 1.23 & 1.19 & 0.80 & 0.56 & 0.76 & 0.68\\
T1 & 3.55 & 1.33 & 1.19 & 0.65 & 0.45 & 0.71 & 0.82\\
T2 & 3.70 & 1.43 & 1.18 & 0.46 & 0.31 & 0.65 & 0.99\\
T3 & 3.82 & 1.55 & 1.18 & 0.25 & 0.16 & 0.59 & 1.19\\
T4 & 3.90 & 1.68 & 1.17 & 0.02 & 0.01 & 0.55 & 1.43\\
T5 & 3.95 & 1.81 & 1.16 & -0.19 & -0.11 & 0.54 & 1.70\\
T6 & 3.98 & 1.96 & 1.16 & -0.35 & -0.19 & 0.59 & 2.02\\
T7 & 4.01 & 2.11 & 1.15 & -0.43 & -0.20 & 0.70 & 2.38\\
T8 & 4.08 & 2.26 & 1.15 & -0.36 & -0.09 & 0.90 & 2.79\\\hline
\end{tabular}
\caption{The $i-z$, $z-Y$, $Y-J$, $J-H$, $H-K$, $K-W1$, $W1-W2$
  template colours for dwarfs ranging from M5 to T8. {\em All
    photometry is on the Vega system.}}
\label{tab:template}
\end{table}
\begin{table*}\footnotesize
\centering
\begin{tabular}{l l l l l l l l}
\hline\hline
{Colour} & {$a_0$} & {$a_1$} & {$a_2$} & {$a_3$} & {$a_4$} & {$a_5$} & {$a_6$}\\\hline
$i-z$ & --9.251 & +3.99519 & --0.540767 & +0.03437326 & --0.001010273 & +1.114655e-05 & 0\\
$z-Y$ & --0.942 & +0.44857 & --0.040668 & +0.00157836 & --1.9718e-05 & 0 & 0\\
$Y-J$ (<L7) & --0.174 & +0.16790 & --0.004615 & 0 & 0 & 0 & 0\\
$Y-J$ (>L8) & +1.312 & --0.00592 & 0 & 0 & 0 & 0 & 0\\
$J-H$ & --2.084 & +1.17016 & --0.199519 & +0.01610708 & --0.000593708 & +7.94462e-06 & 0\\
$H-K$ & --1.237 & +0.69217 & --0.114951 & +0.00946462 & --0.000361246 & +5.00657e-06 & 0\\
$K-W1$ & --4.712 & +2.37847 & --0.444094 & +0.04074163 & --0.001910084 & +4.367540e-05 & --3.83500e-07\\
$W1-W2$ & --0.364 & +0.17264 & --0.015729 & +0.00048514 & 0 & 0 & 0\\\hline
\end{tabular}
\caption{Dependence of colour $c$ on spectral type $t$ defined by
  polynomials $c=\sum_{i=0}^Na_{i}t^{i}$, where the correspondence
  between $t$ and spectral type is $5-9$ represents M5$-$M9, 10$-$19
  represents L0$-$L9, and 20$-$28 represents T0$-$T8. The polynomials
  are valid over M5$-$T8. {\em All photometry is on the Vega system.}}
\label{polynomials}
\end{table*}

\begin{figure} 
\includegraphics[width=8.5cm]{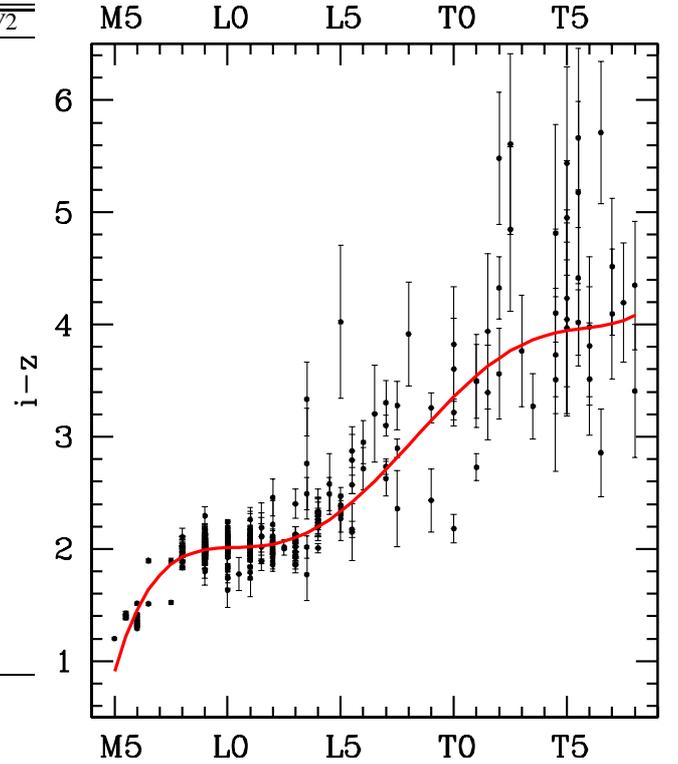} 
\caption{Plot of $i-z$ colour vs. spectral sub-type for MLT dwarfs in the
  UKIDSS LAS DR10 footprint. The error bars plotted show the random
  photometric errors. The fitted curve provides the template colours
  listed in Table \ref{tab:template}. In making the fit, a colour error
  of 0.07 mag. (i.e. 0.05 mag. in each band) is added in quadrature to
  the random photometric errors to account for intrinsic scatter in
  the colours. The very large errors in the T dwarf regime mean that
  the curve is poorly defined in this region, as discussed in the
  text. The vertical scale is the same as in Figure \ref{poly}. The
  outlying blue T0 dwarf is discussed in \S\ref{subsec:accuracy}. {\em All
    photometry is on the Vega system.} }
\label{polyiz}
\end{figure}

\begin{figure} 
\includegraphics[width=8.5cm]{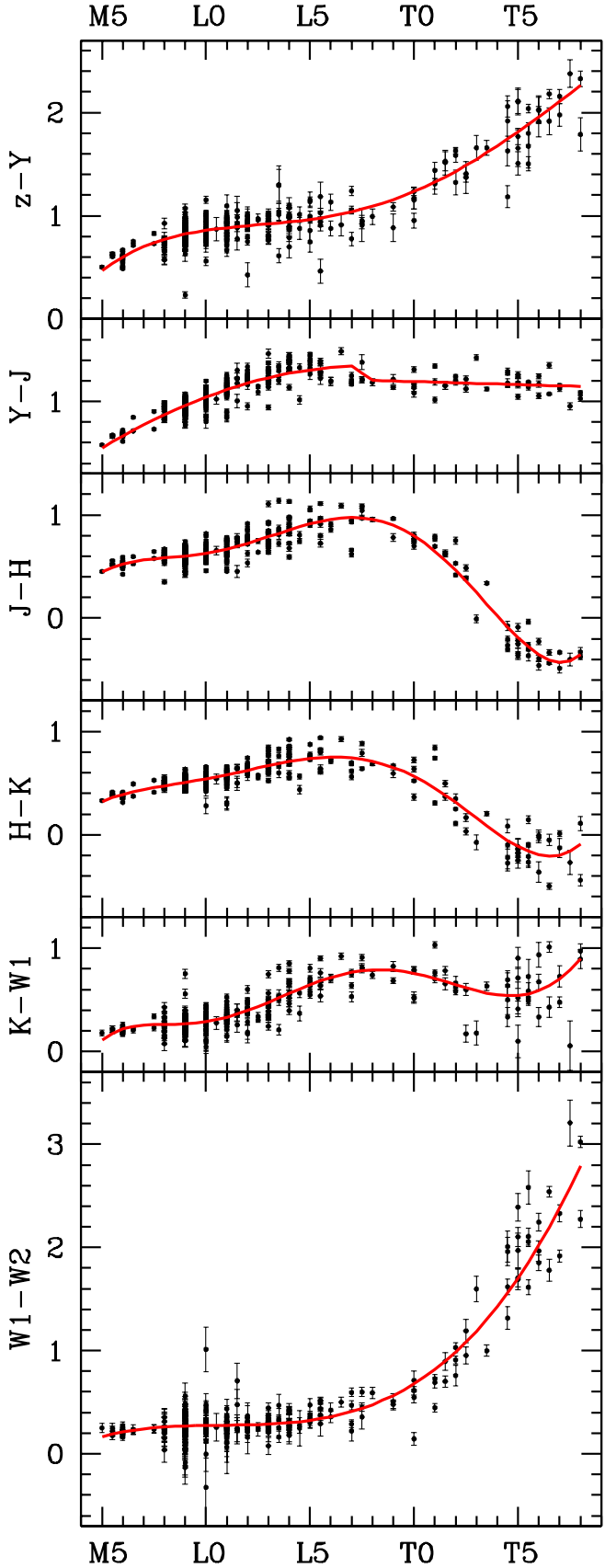} 
\caption{Plot of colours $z-Y$, $Y-J$, $J-H$, $H-K$, $K-W1$, $W1-W2$
  vs. spectral sub-type for MLT dwarfs in the UKIDSS LAS DR10
  footprint. The errorbars plotted show the random photometric
  errors. The fitted curve provides the template colours listed in
  Table \ref{tab:template}. In making the fit, a colour error of 0.07
  mag. (i.e. 0.05 mag. in each band) is added in quadrature to the
  random photometric errors to account for intrinsic scatter in the
  colours. The vertical scale is the same as in Figure
  \ref{polyiz}. {\em All photometry is on the Vega system.} }
\label{poly}
\end{figure}
\begin{figure} 
\includegraphics[width=9.0cm]{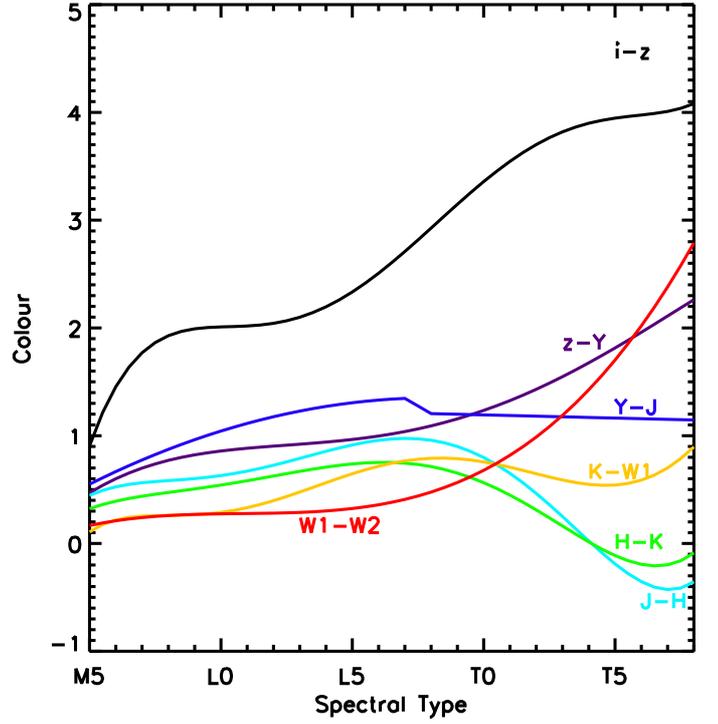} 
\caption{The colour curves from Figures \ref{polyiz} and \ref{poly} 
  plotted in a single
  figure in order to compare the relative usefulness of different
  colours in classifying different spectral types in the LT sequence.}
\label{poly_alltemps}
\end{figure}

\begin{figure} 
\includegraphics[width=8.5cm]{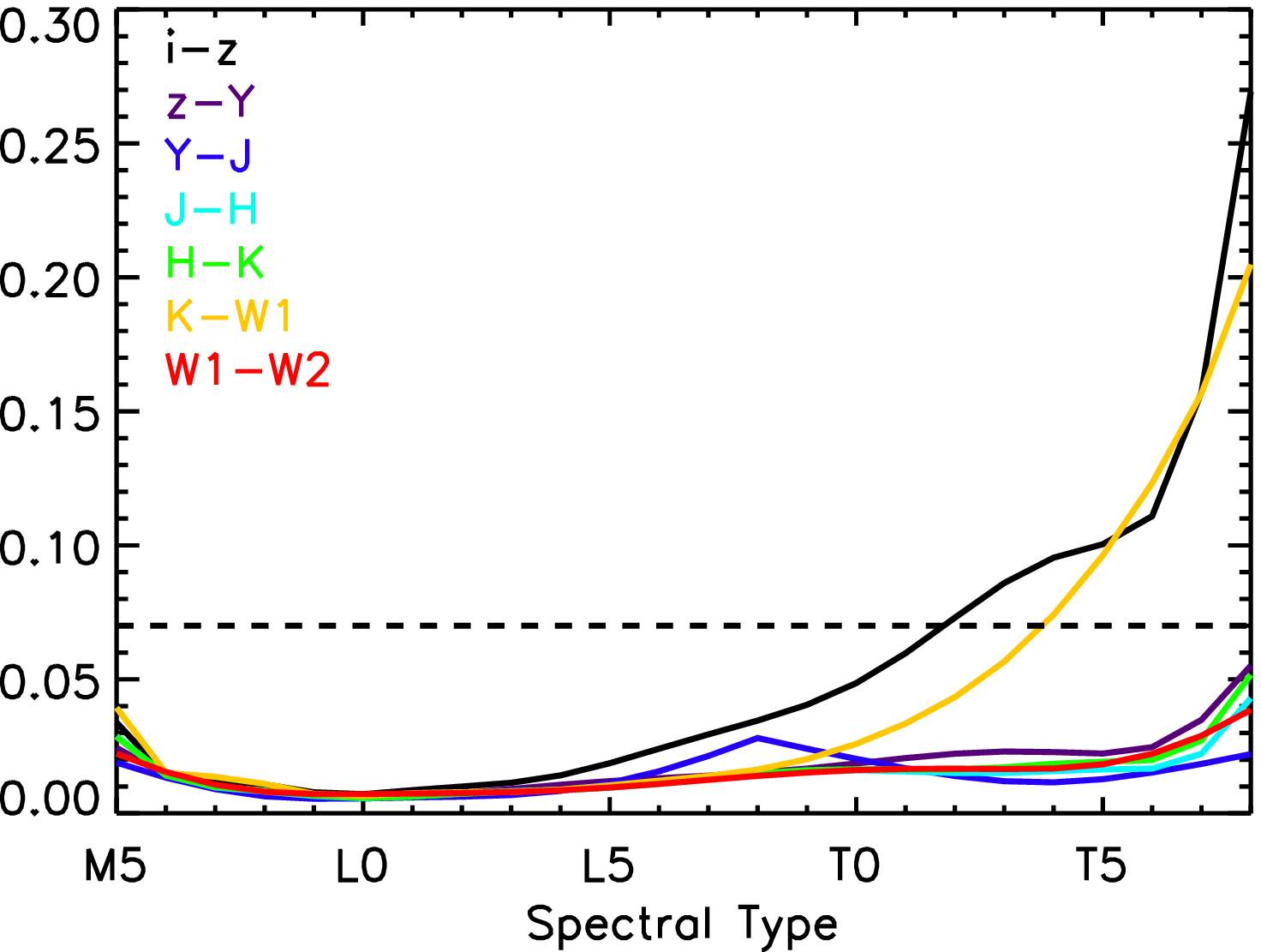} 
\caption{The uncertainties of the fits of the colour polynomials as a function of
  spectral type. With the exception of the $i-z$ and $K-W1$ curves the
  uncertainties are significantly smaller than the intrinsic scatter
  (marked by the dashed line).}
\label{covar_err}
\end{figure}

The templates used in {\em photo-type} fit quasars, white dwarfs,
stars and brown dwarfs. The template colours for quasars were taken
from \citet{quasar}. Templates for quasars with weak, typical, and
strong lines were included, for a range of reddening, $E(B-V)=0.00,
0.10, 0.20, 0.30, 0.40, 0.50$, for the redshift range $0<z<3.4$.  The
template colours for white dwarfs and main-sequence stars earlier than
M5 used in this work are taken from \citet{hewett06}. These are
included for completeness, but in fact are not relevant for objects
with colour $Y-J>0.8$. The star and quasar templates only cover the
bands $izYJHK$ and were used in the initial selection of candidate L
and T dwarfs, before the classifications were refined by including the
WISE colours, as explained in $\S3$.

The templates in \citet{hewett06} for stars of class M5 and later, and
for L and T dwarfs, were found to be inadequate for accurate
classification. These colours were computed from spectra, and before
any $Y$ band data with UKIDSS had been taken. Now that UKIDSS is
complete it is possible to improve on the colour relations of
ultra-cool dwarfs using photometry of known sources within the UKIDSS
footprint.  These will be more accurate than colours computed from
spectra, in particular for the $z$ band where the computed fluxes
are very sensitive to the exact form of the red cutoff of the band,
defined by the declining quantum efficiency of the CCDs.

To compute template colours for ultracool dwarfs we fit polynomials to
colours of known dwarfs as a function of spectral type, as shown in
Figures \ref{polyiz} and \ref{poly}. We searched DwarfArchives.org,
and additional recent published samples (\S\ref{subsec:bias}), for
spectroscopically classified L and T dwarfs within the UKIDSS DR10
footprint. In \S\ref{subsec:photom} we describe the creation of a
catalogue of point sources matched over 3344 deg$^2$ of SDSS, UKIDSS DR10, and WISE,
in the magnitude range $13.0<J<17.5$, with $Y-J>0.8$. Of the sample of
spectroscopically classified L and T dwarfs, 190 appear in our
catalogue i.e. are classified as stellar, have reliable matched
photometry in $izYJHKW1W2$, and meet the magnitude and colour
selection criteria\footnote{The missing sources are discussed in
  \S\ref{subsec:bias}.} Our fundamental assumption is that the measured
  colours of these sources are representative of the colours of the L
  and T populations. Within this sample there are 150 L dwarfs and 40
  T dwarfs. We discuss possible bias of this sample in the next
  section. We supplemented the sample of L and T dwarfs with 111 cool
  M stars of spectral type M5 to M9 from the SDSS spectroscopic
  catalogue \citep{ahn12} in order to tightly constrain the
  polynomials at the M/L boundary.

The 40 T dwarfs are classified on the revised system of \citet{burgasser06b}, based on near-infrared spectroscopy, and so our colour fits are anchored to this system for T dwarfs. Of the 150 L dwarfs, 116 have optical classifications on the system of \citet{kirkpatrick99}, and of these 16 also have near-IR classifications. For these 16 cases we averaged the two classifications quantised to the nearest half spectral sub-type. The remaining L dwarfs have near-IR classifications only. Of the near-IR classifications, close to half use the SpeX prism spectral library \citep{burgasser2014}\footnote{hosted at
 \texttt{http://pono.ucsd.edu/$\sim$adam/browndwarfs/spexprism/}\\ \texttt{html/mldwarfnirstd.html}}, which is anchored to \citet{kirkpatrick99}, while the remainder use the index system of \citet{geballe02}. Since we wish {\em photo-type} to be anchored to the optical classification system we checked for any bias introduced by including the near-IR spectral types. We computed template colours, as described below, for the two cases, including and excluding near-IR spectral types for the L dwarfs. We then classified a large sample of L dwarfs (the 1077 described in \S3) using both sets of templates and looked at the differences in spectral type for the two sets of templates. The result was a negligible offset, $\mu=0.06$ spectral types, and small scatter, $\sigma=0.23$, with no clear trends with spectral type, confirming that for L dwarfs {\em photo-type} is accurately anchored to the optical system of \citet{kirkpatrick99}. 
     
Although the SpeX prism near-infrared spectral library for L dwarfs is tied to the optical classification scheme of \citet{kirkpatrick99}, there exist spectrally peculiar sources where the optical and near-IR classifications differ, in some cases by more than two spectral sub-types. In the same way, for peculiar sources the {\em photo-type} classifications may differ from the optical classifications. One of the virtues of the {\em photo-type} method is that it provides both a best-fit spectral type, and a goodness of fit statistic (the multi-band $\chi^2$) which can be used to identify peculiar sources. In this sense it provides more information than a simple classification based on optical spectroscopy or near-infrared spectroscopy alone.

Denoting colour by $c$ and spectral type by $t$, with numerical values M5$-$M9 as 5$-$9, L0$-$L9 as 10$-$19 and
T0$-$T8 as 20$-$28, we fit polynomials $c=\sum_{i=0}^Na_{i}t^{i}$ via
$\chi^2$ minimisation\footnote{$\chi^2 =
\displaystyle\sum_{i=1}^{N}\Bigg(\frac{c_i-f_i}{\sigma_i}\Bigg)^2$,
where $c$ is the colour of the individual source, and $f$ is the
colour of the polynomial fit}, over the 8 bands.

In the development of this analysis it was immediately clear that the
scatter in the colours is larger than the typical photometric error
i.e. there is an intrinsic scatter in the colours, due e.g. to
variations in metallicity, surface gravity, cloud cover, and unresolved binaries in
the sample, as well as uncertainty in the spectral classification. It
is important to allow for this scatter in fitting the curves, or the
fits could be affected by outlying points with small photometric
errors.

We estimated the intrinsic scatter for each colour in an iterative
fashion as follows. We first guessed the intrinsic scatter by eye, and
added this value in quadrature to the photometric error on each
point. We then found the lowest order polynomial that provided a good
fit to the data. We then re-estimated the intrinsic scatter as the
value that, added in quadrature to the photometric error on each
point, and summed over all points, matched the measured variance about
the fitted polynomial, having identified and removed any discrepant
outliers. Averaging over all the colours we measured an intrinsic
scatter of 0.07 mag, which we adopted as the intrinsic scatter for
all colours. The implementation involved adding 0.05 mag. error
(i.e. $0.07/\sqrt{2}$) in quadrature to the photometric error in each
band for every object. This may be viewed as the uncertainty on the templates.

Having established a suitable value for the intrinsic scatter in the
colours, the polynomials were refit, starting with a linear fit, and
then successively increasing the order of the polynomial, only
provided a significant improvement in the fit was achieved,
$\Delta\chi^2>7$, in order to prevent over-fitting\footnote{At each
  stage, in adding one free parameter, the improvement in $\chi^2$
  will be distributed as the $\chi^2$ distribution with one degree of
  freedom. Then $\Delta\chi^2>7$ corresponds to $>99\%$
  significance.}. In most cases a fourth or fifth order polynomial was
sufficient. The fitted polynomials are shown in Figures \ref{polyiz}
and \ref{poly}. The coefficients of the polynomials are provided in
Table \ref{polynomials}, and the template colours are provided in
Table \ref{tab:template}. 

An additional source of error that should be accounted for in
  classifying sources is the uncertainty in the polynomial fits
  themselves. Since some curves are more tightly constrained by the
  data than others, by incorporating this uncertainty, and its
  variation with spectral type, the different curves are then
  correctly weighted. We have established the uncertainties from the
  covariance matrices of the polynomial fits, and the results are
  plotted in Fig. \ref{covar_err}. We have incorporated these errors
  in classifying sources, and found that in fact they have no
  significant influence on the classifications. The analysis leading
  to this conclusion is presented in \S\ref{accuracy:simulated}. \\

A number of the individual colour plots are discussed below:
\begin{itemize}
\item {\em i-z colour:} The $i-z$ colour polynomial is not well
  defined for T dwarfs. This is because nearly all the T dwarfs are
  very faint in the $i$ band, and so the errors are large, $\sim 0.5$
  mag. or greater. It would be useful to measure accurate $i-z$
  colours for this sample in order to better establish the shape of the curve and
  the intrinsic scatter, but this is difficult because the $z$ band
  used will have to match the SDSS passband very closely (as explained
  above). In practise the difficulty of measuring the $i-z$ curve
  accurately in the T dwarf regime is not critical because, of course,
  most of the candidates are very faint in $i$, and so have large
  photometric errors, meaning that the contribution to the total
  $\chi^2$ from the $i$ band is relatively small. We show later
    (\S\ref{accuracy:simulated}) that including the $i$ band improves the
    accuracy of the classification of L dwarfs, but not of T dwarfs.
 
\item {\em Y-J colour:} We found we were unable to fit the $Y-J$ curve
  satisfactorily with a single polynomial. We attribute this to a
  discontinuity in the relation near spectral type L7. From inspection
  of the SpeX spectra of the near-IR spectral standards the jump
  appears to be associated with the rapid weakening of FeH absorption
  in the $Y$ band between spectral types L6 and L8
  \citep{burgasser02cloud}. To check, and to decide where to impose
  the discontinuity, we computed synthetic colours from the SpeX
  near-IR spectral standards. These are provided in Table
  \ref{spec_standards} (\S\ref{subsec:accuracy}), and show a break in $Y-J$ colour between
  L7 and L8. We therefore fit two separate polynomials to the data,
  with a quadratic for types $\leq$L7, and a linear fit for types
  $\geq$L8, joined by a straight line from L7 to L8. The step between
  L7 and L8 is 0.14 mag. In Paper II we show the same $Y-J$ plot for
  the new sample of L and T dwarfs, and the jump is seen more clearly
  in the larger sample.

\item {\em J-H, J-K colours:} The flattening and return of the colour
  to redder values at the end of the T sequence appears to be real,
  rather than an artefact of the fitting procedure, as it can also be seen
  in the plots in \citet{leggett10} and \citet{cushing11}.

\end{itemize}

Figure \ref{poly_alltemps} illustrates the usefulness
of different colours in the classification of dwarfs of different
spectral type. The steeper the curve, the more accurate the
classification, with the exception of $i-z$ in the T dwarf region
where the errors are larger. Therefore it is evident that spectral classification
will be most accurate in the region from about T1 to T5, because in
this region several of the colour relations are steep, whereas around
L6 several of the colour curves are rather flat and classification
will be less accurate. We quantify the accuracy of classification in
\S\ref{subsec:accuracy}.

\subsubsection{Priors for classification}

We now return to the assumption of flat priors for all the
templates. To check the potential contamination of the sample by
quasars, we used the quasar templates as synthetic sources (adding
photometric errors as appropriate) and picked out the quasar, from the
full set, that provided the best-fit to any dwarf along the sequence
L0 to T8. The best match between the two template sets is a reddened
quasar $E(B-V)=0.5$ of redshift $z=2.7$ matched to a L1.5 dwarf, for
which the goodness of fit is $\chi^2=92$ (for six degrees of freedom).
With such a poor fit it is clear that L and T dwarfs will be easily
discriminated from reddened quasars, independent of priors, within
reason.

\begin{figure} 
\includegraphics[width=8.5cm]{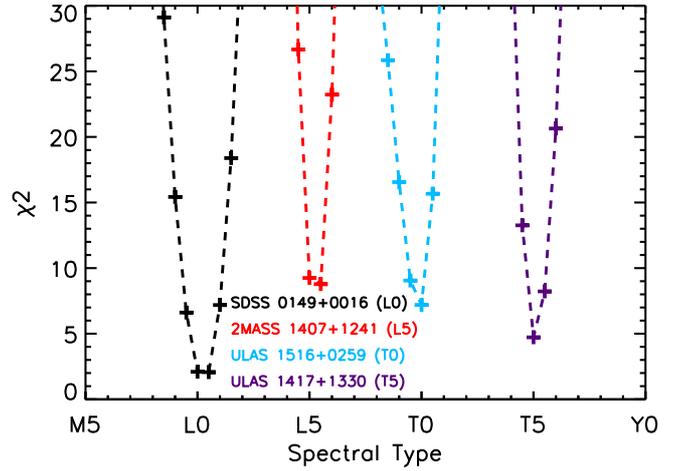} 
\caption{Plot of $\chi^2$ (computed from Equation \ref{chi2_eqn}) against spectral type, in the
  classification of four known L and T dwarfs, as follows: SDSS
  J0149+0016 (L0, $J=17.18\pm0.03$), 2MASS J1407+1241
  (L5, $J=15.33\pm0.004$), ULAS J1516+0259 (T0, $J=16.88\pm0.02$), and
  ULAS J1417+1330 (T5, $J=16.77\pm0.01$). In 
  each case the $\chi^2$ curves show no degeneracies, and are sharp,
  indicating accurate classification (quantified in \S\ref{subsec:accuracy}).}
\label{chisqplots}
\end{figure}

In a flux-limited sample early L dwarfs are much more common than late
L dwarfs. Because of uncertainty in the classifications, the
assumption of flat priors along the LT sequence will lead to a bias in
the classifications (akin to Eddington bias). The extent of the bias
depends on the luminosity function and on the precision of the
classifications i.e. how sharp the curve of $\chi^2$ against spectral
type is. In Figure \ref{chisqplots} we plot $\chi^2$ against spectral
type for four known L and T dwarfs from the sample of 190 previously
known dwarfs. These plots show that the $\chi^2$ minimum is very sharp
and unambiguous, minimising any bias in the classification. The
Eddington bias will be corrected for in computing the luminosity
function in a future paper.

\subsection{Possible sources of bias in the colour relations}
\label{subsec:bias}

In this section we discuss possible bias in the derived colour
relations. This issue is closely related to the question of
completeness, but we defer a detailed discussion of the completeness
of the new sample to Paper II. As stated in \S\ref{subsec:templates}, our
fundamental assumption is that the sample of 190 L and T dwarfs used
to define the polynomial colour relations is representative of the L
and T population i.e. the mean and spread in colour at each spectral
type. We first examine reasons for missing a few of the catalogued
objects.

We searched DwarfArchives (update of 6th of November 2012) for known L
and T dwarfs with $13.0<J<17.5$ in the UKIDSS LAS $YJHK$ footprint,
supplemented by the recent samples of \citet{burningham13},
\citet{DayJones13}, \citet{kirkpatrick11}, and \citet{mace13}. A few
sources with unreliable photometry in any of the eight bands
(e.g. blended with a diffraction spike, or landing on a bad CCD row)
were discarded. In addition, three sources were classified as stellar
in 2MASS, but appear elongated in UKIDSS because they are
binaries. This means there is a small bias against finding binaries
where the angular separation of the pair is a few tenths of an arcsec,
but this should not have any significant effect on the colour
relations. The final sample comprises 192 known L and T dwarfs in the
surveyed area, with reliable photometry, and classified as stellar. \\

A further two sources are missing from our sample:
\begin{itemize}
\item WISEPC J092906.77+040957.9: This is a T6.5 dwarf discovered by
\citet{kirkpatrick11}. The source was missed because its motion
between the epochs of the $Y$ and $K$ UKIDSS observations was
$2\farcs06$, which is just greater than the $2\farcs0$ UKIDSS
internal matching criterion.

\item SDSS J074656.83+251019.0: This source is an L0 dwarf discovered by
\cite{schmidt10}. It has $Y=17.37$, and $J=16.58$, giving
$Y-J=0.79$, meaning it is just bluer than our selection limit
$Y-J=0.8$. A second epoch $J$ measurement gave $J=16.60$. The source is
apparently anomalously blue in $Y-J$, as the typical colour of an L0
dwarf is $Y-J=1.04$ (Table \ref{tab:template}).
\end{itemize}
The inclusion of these two sources would not significantly affect the
colour relations. Therefore whether the colour polynomials are biased
is a question of whether the sample in DwarfArchives contains
significant colour selection biases, or whether there are any
significant populations of L and T dwarfs with peculiar colours that
remain undiscovered.

The samples of \citet{kirkpatrick11} and \citet{mace13}, contain many
late T dwarfs selected with WISE. To check specifically the $W1-W2$ relation we
created a comparison sample after excluding those sources already used
in the curve fitting (i.e. those in the UKIDSS LAS footprint).  We
compared the $W1-W2$ polynomial obtained for this catalogue to our
template polynomial, finding it to be identical to within 0.01
mag. over the full LT range.

We also re-examined the bias, noted by \citet{schmidt10}, that early L
dwarfs in DwarfArchives (at the time) were on average redder in $J-K$
by $\sim 0.1$ mag than those in their much larger sample.
\citet{schmidt10} selected sources for spectroscopy with a
sufficiently blue cut in $i-z$ to ensure they included all L dwarfs,
and so their $J-K$ colours should be unbiased. This large sample is
now included in DwarfArchives, so any bias is much
reduced. \citet{schmidt10} ascribed the bias in the older
DwarfArchives sample to the colour cut $J-K>1.0$ applied by
\citet{cruz03} in selecting L dwarfs, meaning that bluer L dwarfs were
excluded. The extent of the bias introduced by a colour cut would
depend on both the intrinsic spread in $J-K$ colour, as well as the
random errors. Our own analysis, which uses the much more precise
UKIDSS photometry rather than the original 2MASS photometry, finds a
smaller bias between these two samples of $0.05$ mag over the spectral
type range L0 to L4. Our analysis suggests that much of the original
bias noted was due to random photometric errors, rather than true
colour bias. Accounting for the relative sizes of the two samples used
in the analysis, any remaining bias in our $J-K$ colour relation would
be $\lesssim0.01$ mag. 

It remains to consider the possibility that significant populations
with substantially different colour relations are underrepresented in
DwarfArchives. These might include unusually blue objects, such as
SDSS J074656.83+251019.0 (discussed above), or unusually red objects
(\citealt{faherty13}, \citealt{liu13}). Such a population would have a
significant influence, e.g. a shift in the colour relations of
$>0.03$mag., if, for example, the population comprised, say, $10\%$ of
the entire L and T population, and their colours were unusual by
$0.3$mag. There is no indication in the polynomial plots (Figs
\ref{polyiz} and \ref{poly}) of any outlying clouds of points that
might hint at such a population. But given the large differences
between the L and T population and quasars (\S\ref{subsec:templates}),
{\em photo-type} should pick up unusual L and T dwarfs, which will be
identifiable in the new sample. We reconsider this question in Paper
II.

\subsection{Unresolved binaries}
\label{subsec:binaries}

A small number of sources are classified as L or T but with large
$\chi^2$. These objects may be peculiar single sources, or could be
unresolved binaries. To check whether any might be unresolved
binaries, we created template colours for all possible L+T binary
combinations, over the range of spectral types L0 to T8. We use the
relation between absolute magnitude in the $J$ band and spectral type
from \citet{dupuy12} to provide the relative scaling of the two
templates, hereafter referred to as S1 and S2. Then for sources with
large $\chi^2$, we compare the improvement in $\chi^2$ achieved by
introducing the extra degree of freedom of a binary fit. Objects with
$\Delta\chi^2>7$ (where $\Delta\chi^2 = \chi^2_{\textrm{single}} - \chi^2_{\textrm{binary}}$) are accepted as candidate binaries.

\begin{figure} 
\includegraphics[width=8.5cm]{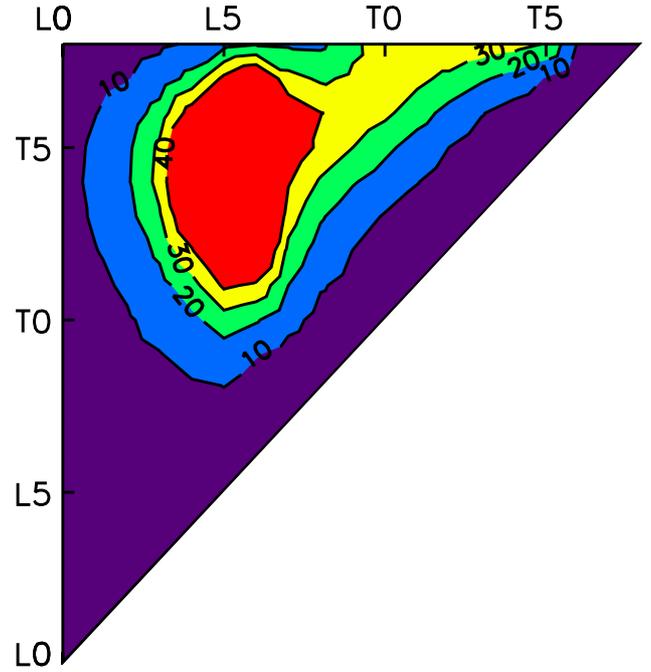} 
\caption{Contour plot illustrating sensitivity to detection of
  unresolved binaries. The axes correspond to the two components of a
  binary. The contours plot the improvement in $\chi^2$ of a binary dwarf
  solution over the best single dwarf solution.}
\label{binarysingle}
\end{figure}

The search for candidate unresolved binaries is effective only over
particular regions of the S1, S2 parameter space because,
given the colour uncertainties, some binary combinations S1+S2 are
satisfactorily fit by the colours of a single source. For example, the
colour template for the combination L0+T5 is very similar to the
colour template of a L0 dwarf. This issue is illustrated in Figure
\ref{binarysingle}.  Here we tested the improvement of a binary fit
over a single fit for different combinations S1+S2. Each point in the
grid represents a binary S1+S2. We created synthetic template colours
for the binary, adding random and systematic errors as before. Then
for each artificial binary we found the best fit single solution and
the best fit binary solution, and recorded the improvement in
$\chi^2$. The contours plot the average improvement in $\chi^2$
achieved with the binary fit, and therefore map out regions where {\em
  photo-type} is sensitive to the detection of binaries.

In Paper II we present spectra of some sources identified as candidate
binaries using {\em photo-type}.

\section{Application}
\label{sec:application}

In this section we describe the creation of a catalogue of point sources
matched across 3344 deg$^2$ of SDSS, UKIDSS and WISE that we search for L and T dwarfs. We also quantify the accuracy of the {\em photo-type} method.

\subsection{Photometric data}
\label{subsec:photom}
 
Our study is concerned with field (as opposed to cluster) brown dwarfs
and uses survey data at high Galactic latitudes. The starting point
for the search is the Data Release 10 (DR10) version of the UKIDSS
Large Area Survey (LAS) which provides photometry in the {\em YJHK}
bands. Point sources are matched to the SDSS and ALLWISE catalogues,
to add, respectively, {\em iz} and {\em W1W2} photometry. Sources are
then classified using the full {\em izYJHKW1W2} data set. The
footprint of the survey is defined by the 3344 deg$^2$ area of the 
UKIDSS LAS, where all four of the {\em YJHK} bands have been observed,
contained within the SDSS DR9 footprint. All sources also possess WISE
photometry from the ALLWISE catalogue.

Creating a matched catalogue requires careful consideration of the
image quality, pixel scale, and flux limits of the different
surveys. The pixel scales of UKIDSS and SDSS are the same, $0\farcs4$,
and the two surveys are well matched in terms of image quality: the
typical seeing in UKIDSS is $0\farcs8$ \citep{warren07}, and in SDSS
is $1\farcs4$ \citep{adelman07}. The WISE raw images on the other hand
have large pixels, $2\farcs75$, and the full width at half maximum
of the {\em W1} and {\em W2} point spread function (PSF) is
$6\arcsec$. The larger PSF results in significant blending of images,
leading to incorrect photometry. Such cases are identified by visual
inspection of the spectral energy distributions and the images. In the
final catalogue, the $7\%$ of sources blended in WISE are classified
using only the 6-band {\em izYJHK} photometry.

We chose the UKIDSS $J$ band for defining the flux range of the
survey, and selected $J=13.0$ as the bright limit, to avoid saturation
in any band. The faint limit of $J=17.5$ was chosen from detailed
consideration of the SEDs of L and T dwarfs, and the detection limits
in each band, as described below. It corresponds to the maximum depth
at which a complete sample with accurate classifications can be
defined, given the data. The average $5\sigma$ detection limit in $J$
is $19.6$, so at $J=17.5$ the S/N of sources is about 35. Using the
template colours for the L and T sequence
($\S\ref{subsec:templates}$), we created synthetic catalogues, with
appropriate photometric errors, for different possible $J$ flux
limits, and compared the synthetic photometry against the detection
limits in each of the other 7 bands. Moving progressively fainter in
$J$, the first objects to fall below any of the detection limits are a
small fraction of the coolest T dwarfs, by $J=16.5$, absent from SDSS
(i.e. fainter than the detection limit in both $i$ and $z$). Therefore
to extend the depth of the survey we implemented a procedure to
include sources undetected in SDSS. 

The difficulty here is
  compounded by the fact that T dwarfs, being nearby, can have
  significant proper motions. Given the epoch differences between the
  optical and near-infrared observations, typically a few years, this requires a search
  radius of several arcsec. Then, in some cases, the nearest SDSS
  match to the UKIDSS target may be the wrong source. The procedure we
  adopted was to find all SDSS sources within a large search radius of
  $10\arcsec$ about the UKIDSS source. Starting with the SDSS match nearest to the target we
  checked whether there was a different UKIDSS source closer to the
  SDSS source, and if so eliminated the SDSS source as a match to our
  target, and proceeded to the next nearest match.  Sources that were
  not matched by this procedure to any source within the $10\arcsec$
  search radius were retained. This allowed us to extend the depth of
the survey to $J=17.5$. At $J=17.5$ some sources fall below the
detection limit in other bands. The incompleteness due to this is very
small and is quantified below.

In detail, the matching procedure adopted was as follows. From the
UKIDSS LAS database we selected objects detected in the full {\em
  YJHK} set, classified as stellar, using {\tt
  -4<mergedclassstat\footnote {The parameter {\tt mergedclassstat}
  measures the degree to which the radial profile of the image
  resembles that of a star, quantified by the number of standard
  deviations from the peak of the distribution.}<4}.  We also eliminated
sources with questionable photometry using the quality flag {\tt
  pperrbits<255} in each band. Sources are matched to SDSS as
described above. The $izYJHK$ catalogue, $13.0<J<17.5$, contains
6\,775\,168 sources. The procedure to match to WISE is cumbersome, so
we reduced the size of the catalogue before matching to WISE, while
retaining all the L and T dwarfs, as follows. L and T dwarfs are
redder in $Y-J$ than all main-sequence stars. So to find L and T
dwarfs we can make the problem more manageable by taking a cut in
$Y-J$. In \S\ref{subsec:templates} we show that the bluest L or T
dwarf template, for type L0, has colour $Y-J=1.04$, and that the
intrinsic scatter in colours is $\sim 0.07$.  On this basis we applied
a cut at $Y-J>0.8$, to produce a sample of 9487 sources. From this
sample we then visually checked all sources classified as missing in
SDSS, eliminating obvious errors (due e.g. to blending or bad
rows). In a small number of cases, where the undetected object is just
visible, we undertook aperture photometry of the SDSS images using the
Image Reduction and Analysis Facility (IRAF;
\citealp{iraf})\footnote{In the final catalogue of 1157 L and T
  dwarfs, there are 10 sources without photometry in both $i$ and $z$,
  that were classified using the other bands.}.

The sample at this stage is dominated by late M dwarfs. To reduce the
sample size further, before matching to WISE, we made a first-pass
classification, using the method described in
\S\ref{subsec:classification}, from the combined $izYJHK$ photometry,
and then limited our attention to the 7503 sources classified as cool dwarfs,
with classification M6 or later. All these sources were then matched
to WISE, using a $10\arcsec$ search radius, to extract the $W1$ and
$W2$ photometry. Only one source, classified by {\em photo-type} as
L1.5\footnote{ULAS J211045.61+000557.0}, was unmatched to WISE,
because it is just below the detection limit.  It is retained,
together with its UKIDSS+SDSS classification. The final, refined,
classifications were obtained from the full $izYJHKW1W2$
photometry. All candidate L and T dwarfs were visually inspected in
all bands. We also plotted up the multiwavelength photometry to
identify likely spurious data.

Finally, we quantified any incompleteness in the original UKIDSS
$YJHK$ catalogue due to objects falling below the detection limit in
any band. Because of their blue $J-K$ colours, at $J=17.5$ late T
dwarfs have $K\sim 18$, and a few sources in the
tail of the random photometric error distribution might be scattered
fainter than the detection limit in $K$. We undertook a full
simulation to quantify the incompleteness, accounting for the UKIDSS
detection algorithm, random photometric errors, the intrinsic spread
in $J-K$ colour (\S\ref{subsec:templates}), and the variation in detection limit across
the survey. The result is that, integrating over the volume of the
survey, the incompleteness is $<0.2\%$ for all spectral types, except
T6 and T7 where the incompleteness is $0.6\%$ and $1.2\%$,
respectively.

In summary, from an area of 3344 deg$^2$ we have selected all
  stellar sources $13.0<J<17.5$ detected in $YJHK$ in the UKIDSS LAS,
  and produced a catalogue of 7503 sources with $Y-J>0.8$, matched to
  SDSS and WISE, that are classified as cool dwarfs, M6 or later, from
  their $izYJHK$ photometry. This is the starting catalogue for a
  search for L and T dwarfs. A handful of sources are undetected in
SDSS or WISE, but these are included in the catalogue. Incompleteness
due to sources not detected in any of the $YHK$ bands is negligible.

Classification by {\em photo-type} using the full $izYJHKW1W2$ photometry
produced a sample of 1157 L and T dwarfs.
This sample is described in Paper II.

Of the 190 sources used in the polynomial fitting, all but one are
successfully classified by {\em photo-type} as ultra-cool dwarfs. The
exception is the unusual L2 dwarf 2MASS J01262109+1428057 discovered
by \citet{metchev08}, which was better fit by a reddened quasar
template. Therefore, of the 192 known L and T dwarfs with reliable
photometry in the surveyed area and magnitude range, 189 are recovered
by our selection and classification method.

\subsection{Classification accuracy}
\label{subsec:accuracy}

In this subsection we estimate the accuracy of {\em photo-type}. To
recap, for a source measured in $izYJHKW1W2$, an intrinsic uncertainty
of 0.05 magnitude is added in quadrature to the photometric error in
each band, and then the $\chi^2$ of the fit to each template is
measured, using Equations \ref{weighted_mean} and \ref{chi2_eqn} from
\S\ref{subsec:classification}. The template providing the minimum
$\chi^2$ fit provides the classification. We estimate the accuracy of
the classification in three ways.
\begin{itemize}
\item In the first case we see how accurately we recover the
spectroscopic classifications of the 189 sources from DwarfArchives.
\item Because this sample is heterogeneous, with classifications based
on spectra covering different wavelength ranges, we also obtained our
own follow-up spectra of candidates from Paper II, to obtain a second
assessment of the classification accuracy, from a homogeneous
spectroscopic sample.
\item The third estimate uses Monte Carlo methods to create artificial
catalogues of colours of all spectral types, over the magnitude range
of the catalogue, to estimate the classification accuracy as a
function of spectral type and magnitude. We also investigate by how much the
accuracy degrades as different bands are removed, to quantify the
usefulness of those bands.
\end{itemize} There is good agreement between the various methods
discussed in \S\ref{accuracy:DA}, \S\ref{accuracy:spex} and
\S\ref{accuracy:simulated} that classification by {\em photo-type} is
accurate to a root mean square error ({\em r.m.s.}) of one spectral sub-type 
over the magnitude range of the sample.

\subsubsection{Comparison against known sources from DwarfArchives}
\label{accuracy:DA}
\begin{figure} 
\includegraphics[width=9.0cm]{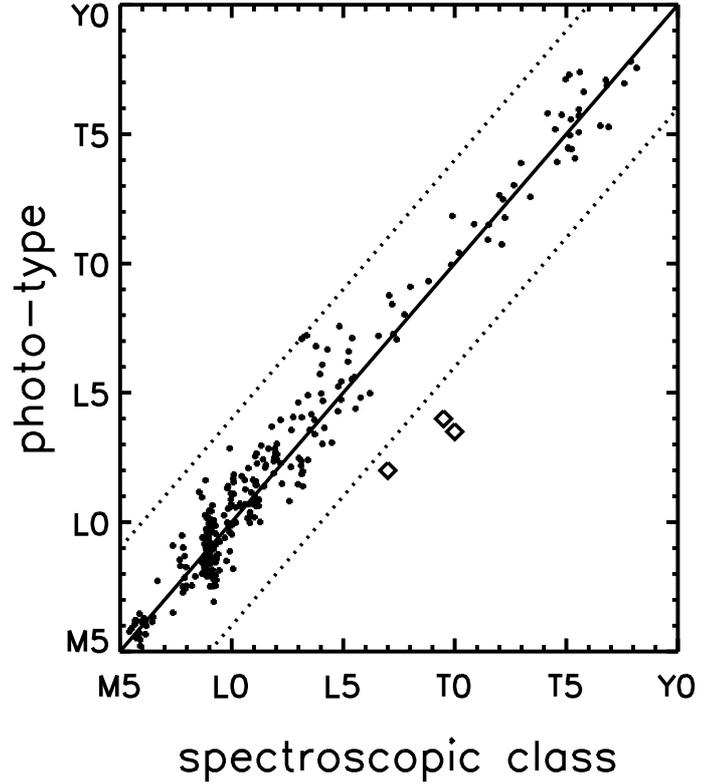} 
\caption{Comparison of {\em photo-type} classification
  vs. spectroscopic classification for the 189 known L and T dwarfs
  from DwarfArchives, as well as 111 late M stars. Classifications
  were computed to the nearest half spectral sub-type. Because the
  classifications are quantised, small offsets have been added
  in order to show all the points. The dashed lines mark
  misclassification by four spectral sub-classes. The three outliers
  marked by diamonds are discussed in the text.}
\label{chi_spec_wise} 
\end{figure}

In Figure \ref{chi_spec_wise} we plot the {\em photo-type}
classification against the spectroscopic classification for the 189 L
and T dwarfs from DwarfArchives (i.e. excluding the single
misclassified source), together with 111 M stars. The vertical scatter
in this plot is a measure of the accuracy of the classification. There
is a contribution to the variance from the quantisation of the
spectral classification. Therefore for this plot the {\em photo-type}
classifications were measured to the nearest half sub-type, by
interpolating the colours in Table \ref{tab:template}. This reduces
the contribution of quantisation to the variance, in order to be able
to measure the scatter accurately. To account for outliers we estimate
the vertical scatter in this plot using the robust estimator
\begin{equation}
\sigma=\frac{{\sum^N_{i=1} |\Delta{t}|}}{N}\frac{\sqrt{2\pi}}{2}
\label{robust}
\end{equation}
where $\Delta{t}$ is the difference between the {\em photo-type}
classification and the spectroscopic classification. The three
outliers marked are discussed below. For the L and T dwarfs we
measure, respectively, $\sigma_{\textrm{L}}=1.5$ and
$\sigma_{\textrm{T}}=1.2$. This may be
considered an upper limit to the uncertainty in the type because there
is a contribution to the scatter from the spectroscopic classification
itself. Some of this scatter comes from the fact that spectroscopic
classification is based on a restricted portion of the photometric
wavelength range covered in this study ($0.75-4.6\mu$m). A 
peculiar source might have a different spectral sub-type if
classified in the optical or the near-infrared. The {\em photo-type}
method will smooth out such differences because of the broad
wavelength coverage. This discussion suggests that $\sigma=1$ is a reasonable
assessment of the accuracy of {\em photo-type}.  There is an element
of circularity in using the same L and T dwarfs used to define the
colour templates in measuring the classification accuracy. In
principle this should not be a concern, since the number of
objects used is very much greater than the number of parameters in the
fitting. Nevertheless it motivates checking the classification
accuracy by other means.

There are three sources in the plot, marked by diamonds, where the 
{\em photo-type} classifications differ from the spectroscopic
classifications by more than four sub-types. These outliers are now
discussed in detail.

\begin{figure} 
\includegraphics[width=9.0cm]{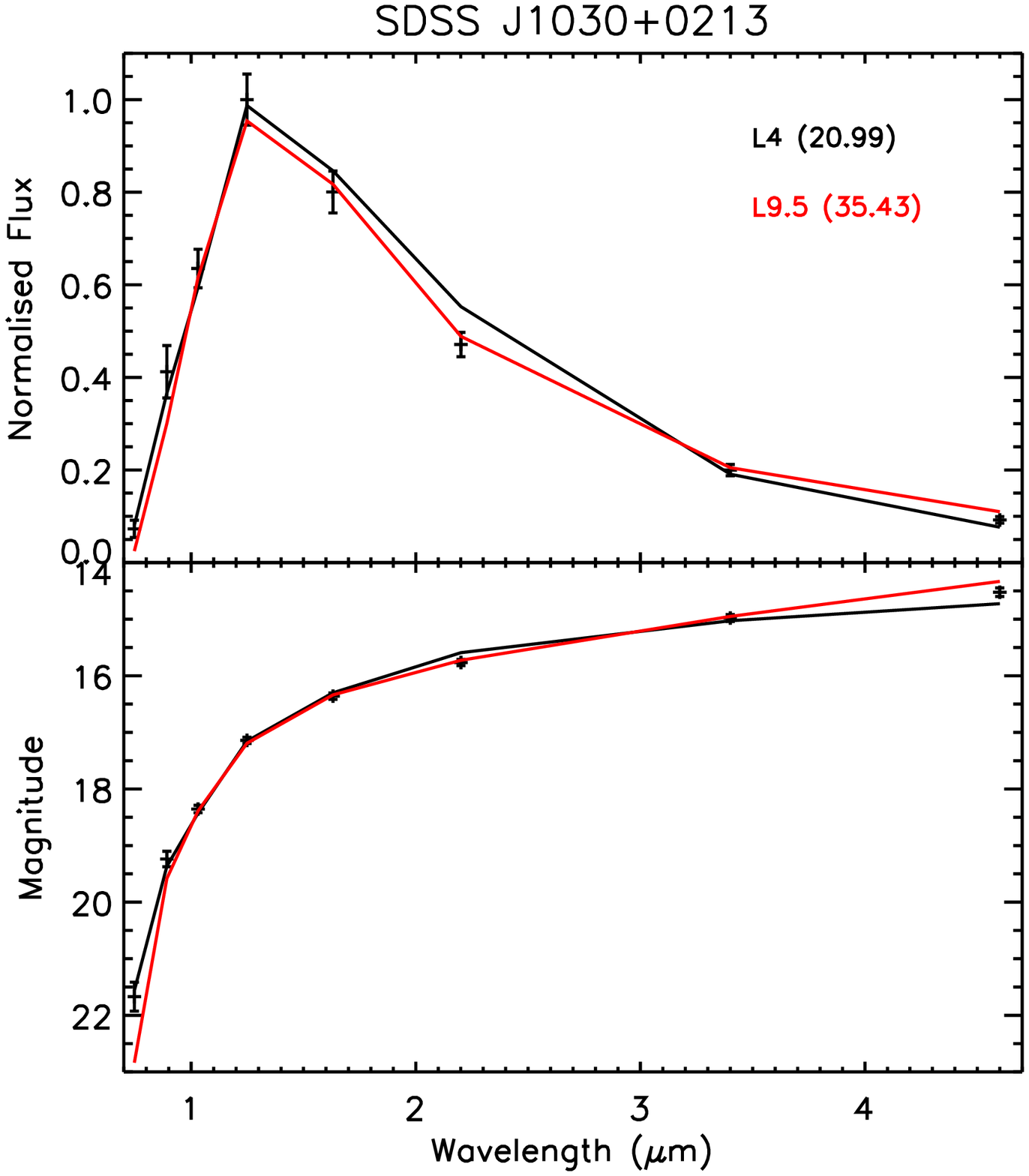} 
\caption{SED of the source SDSS J1030+0213 compared to the {\em
    photo-type} classification of L4 (black line), and the spectroscopic
  classification of L9.5 (red line). The upper plot uses flux ($f_\lambda$),
  normalised to J, and the lower plot uses magnitudes.}
\label{L4orL9} 
\end{figure}

\begin{figure} 
\includegraphics[width=9.0cm]{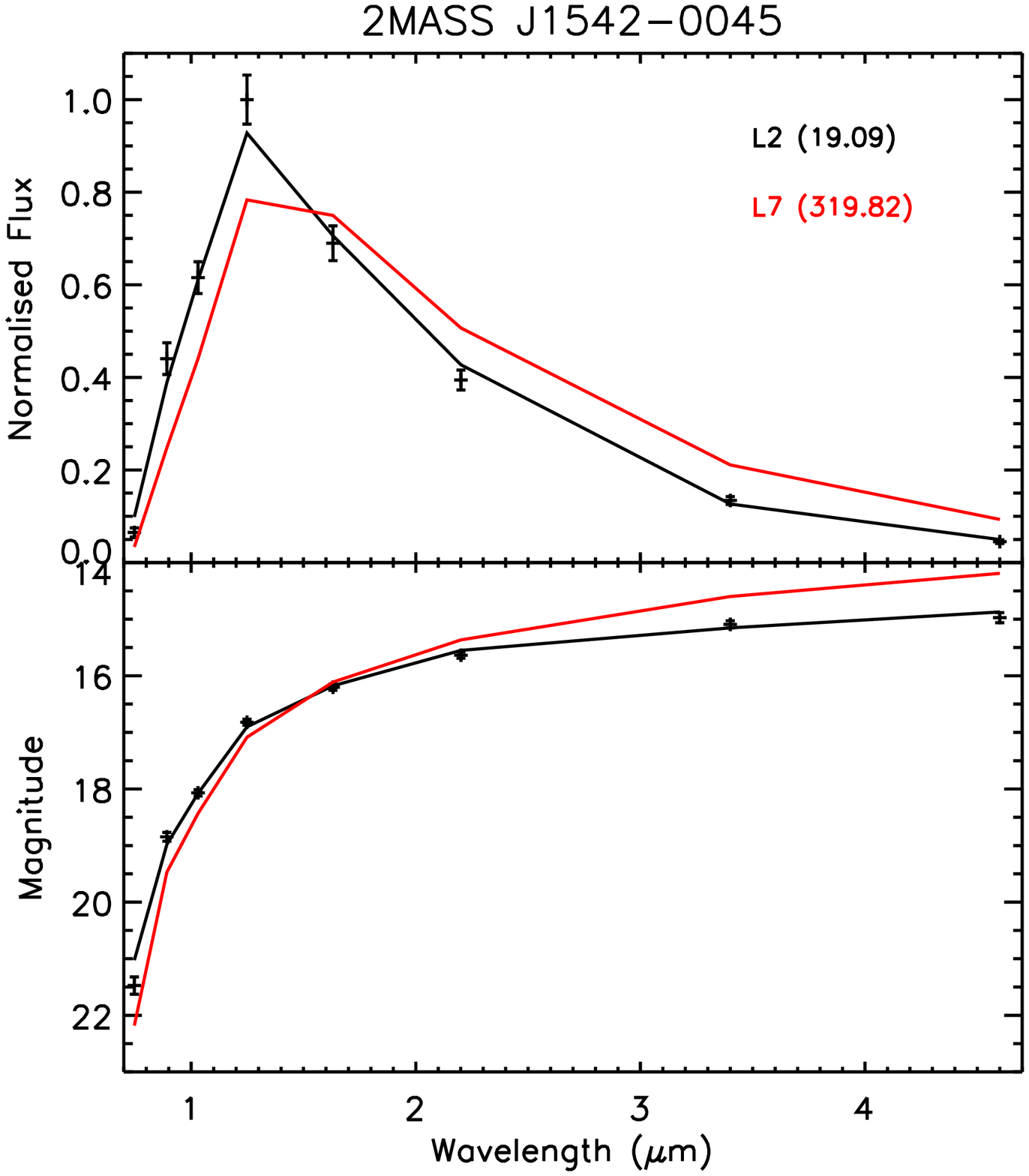} 
\caption{SED of the source 2MASS J1542-0045 compared to the
  {\em photo-type} classification of L2 (black line), and the spectroscopic
  classification of L7 (red line). The upper plot uses flux ($f_\lambda$),
  normalised to J, and the lower plot uses magnitudes.}
\label{L2orL7} 
\end{figure}

\begin{figure} 
\includegraphics[width=9.0cm]{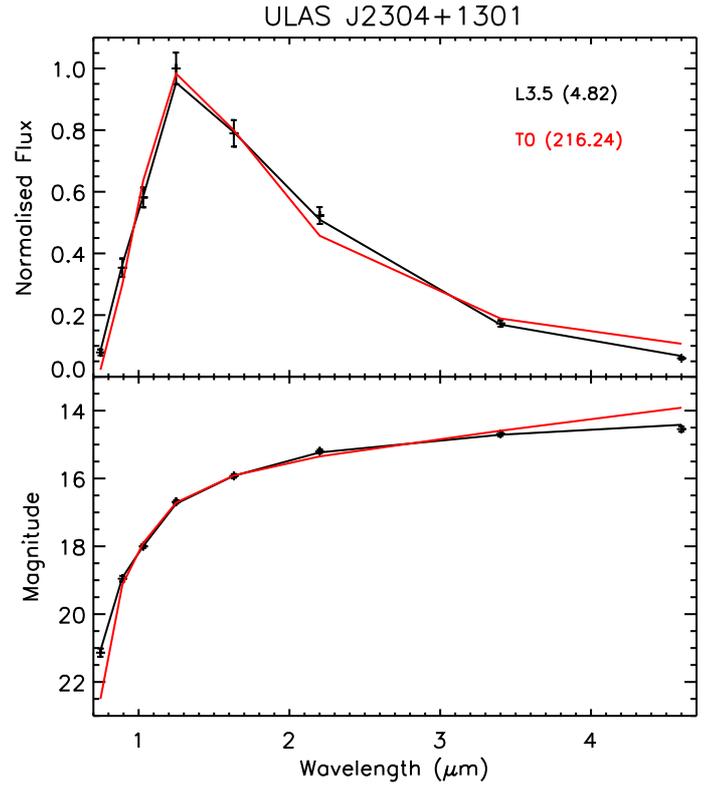} 
\caption{SED of the source ULAS J2304+1301 compared to the
  {\em photo-type} classification of L3.5 (black line), and the spectroscopic
  classification of T0 (red line). The upper plot uses flux ($f_\lambda$),
  normalised to J, and the lower plot uses magnitudes.}
\label{L3.5orT0} 
\end{figure}

\begin{table*}[ht]\footnotesize
\begin{tabular}{l l l l l l}
\hline\hline
{Object Name} & {2MASS Designation} & {SpT} & {$Y-J$} & {Res} & {Reference}\\\hline
VB 8 & J16553529--0823401 & M7 & 0.72 & 120 & \citet{burgasser08}\\
VB 10 & J19165762+0509021 & M8 & 0.81 & 120 & \citet{burgasser04b}\\
LHS2924 & J14284323+3310391 & M9 & 0.97 & 120 & \citet{burgasser06c}\\
2MASP J0345+2540 & J03454316+2540233 & L0 & --  & 75 & \citet{burgasser06c}\\
2MASSW J2130-0845 & J21304464--0845205 & L1 & 1.09 & 120 & \citet{kirkpatrick10}\\
Kelu-1 & J13054019--2541059 & L2 & 1.24 & 120 & \citet{burgasser07b}\\
2MASSW J1506+1321 & J15065441+1321060 & L3 & 1.30 & 120 & \citet{burgasser07a}\\
2MASS J2158-1550 & J21580457--1550098 & L4 & 1.30 & 120 & \citet{kirkpatrick10}\\
SDSS J0835+1953 & J08350616+1953044 & L5 & 1.29 & 120 & \citet{chiu06}\\
2MASSI J1010-0406 & J10101480--0406499 & L6 & 1.27 & 120 & \citet{reid06}\\
2MASSI J0103+1935 & J01033203+1935361 & L7 & 1.30 & 120 & \citet{cruz04}\\
2MASSW J1632+1904 & J16322911+1904407 & L8 & 1.15 & 75 & \citet{burgasser07a}\\
DENIS-P J0255-4700 & J02550357--4700509 & L9 & 1.15 & 120 & \citet{burgasser06b}\\
SDSS J1207+0244 & J12074717+0244248 & T0 & 1.12 & 120 & \citet{looper07}\\
SDSS J0151+1244 & J01514169+1244296 & T1 & 1.08 & 120 & \citet{burgasser04b}\\
SDSSp J1254-0122 & J12545390--0122474 & T2 & 1.14 & 120 & \citet{burgasser04b}\\
2MASS J1209-1004 & J12095613--1004008 & T3 & 1.07 & 120 & \citet{burgasser04b}\\
2MASSI J2254+3123 & J22541892+3123498 & T4 & 1.23 & 120 & \citet{burgasser04b}\\
2MASS J1503+2525 & J15031961+2525196 & T5 & 1.12 & 120 & \citet{burgasser04b}\\
SDSSp J1624+0029 & J16241437+0029156 & T6 & 1.15 & 120 & \citet{burgasser06d}\\
2MASSI J0727+1710 & J07271824+1710012 & T7 & 1.04 & 120 & \citet{burgasser06d}\\
2MASSI J0415-0935 & J04151954--0935066 & T8 & 1.03 & 120 & \citet{burgasser04b}\\\hline
\end{tabular}
\caption[List of SpeX Prism Spectral Library `standards']{The spectral
  standard templates from the SpeX Prism Spectral Library that are
  used for spectral classification. The $Y-J$ colours computed from
  the spectra are listed.}
\label{spec_standards}
\end{table*}

\begin{table*}[ht]\footnotesize
\begin{tabular}{l l l l l l l l l l}
\hline\hline
{R.A. (2000)} & {decl. (2000)} & {Date (UT)} & {$J\pm J_{err}$ (mag)} &{PhT}&{$\chi^2$}&{SpT} & {Exp. time (sec)} & {Slit Width ($\arcsec$)} & {A0 star}\\\hline
00 26 40.46 & +06 32 15.1 & 29/10/2013 & 14.42$\pm$0.01 & L0.5 & 7.15 & L1 &  $\,\,\,$720 & 0.5 & HD 6457\\
01 09 07.42 & +06 25 59.0 & 29/10/2013 & 14.49$\pm$0.01 & L1.5 & 7.05
& L1 &  $\,\,\,$960 & 0.5 & HD 6457\\
02 46 10.23 & +01 56 44.3 & 20/11/2013 & 16.85$\pm$0.02 & L8 & 10.62 & L8 & 1500 & 0.8 & HD 18571\\
07 41 04.39 & +23 16 37.6  & 20/11/2013 & 16.04$\pm$0.01 & L1 & 2.49 & L1 & 1500 & 0.8 & HD 58296$^1$\\
08 49 37.09 & +27 39 26.8 & 20/11/2013 & 16.37$\pm$0.01 & L2 & 0.77 & L1 & 1440 & 0.8 & HD 71906$^2$\\
09 15 44.13 & +05 31 04.0 & 22/11/2013 & 16.93$\pm$0.01 & T3 & 11.29 & T3 & 1440 & 0.8 & HD 79108$^3$\\
10 29 35.20 & +06 20 28.6 & 20/11/2013 & 16.71$\pm$0.01 & L9 & 12.94 & L9.5 & 1500 & 0.8 & HD 71908\\
10 53 20.24 & +04 52 22.3 & 22/11/2013 & 14.78$\pm$0.01 & L0 & 3.88 & M9 &  $\,\,\,$360 & 0.8 & HD 92245\\\hline
\end{tabular}
\\\scriptsize{$^1$ No WISE data, therefore the $\chi^2$ uses only the $
    izYJHK$ bands \\ $^2$ Low S/N, therefore there is an error of
  $\pm$1 in spectral type \\ $^3$ No WISE data, therefore the $\chi^2$ 
  uses only the $izYJHK$ bands}

\caption{Details of the eight sources observed with SpeX.}
\label{good_sources}
\end{table*}

\begin{itemize}
\item SDSS J1030+0213 was discovered by \citet{knapp04}, and
  classified as L9.5$\pm$1.0 from a near-infrared spectrum. Its type
  varies somewhat depending on which spectral indices are used: <T0
  (H$_2$O in $J$), T1 (H$_2$O in $H$), T0.5 (CH$_4$ in $H$) and L8
  (CH$_4$ in $K$). Using the full 8-band photometry {\em photo-type}
  provides a classification of L4 with $\chi^2 = 20.99$, i.e. a
  relatively poor fit, whereas the L9.5 template has $\chi^2 =
  35.43$. The two fits are illustrated in Figure \ref{L4orL9}. The
  object has $i-Y=3.32\pm0.25$, which is unusually blue compared to
  the template colour $i-Y=4.45$ for L9.5. Neither model fits the $W2$
  data satisfactorily. \citet{radigan2014} found that objects around
  the L/T transition can show significant variability, which provides
  a possible explanation for the poor fits, since the various
  photometric data were taken at different dates.

\item 2MASS J1542-0045 was discovered by \citet{geissler11}, who
  classified it as a peculiarly blue L7, from a near-infrared
  spectrum. Our best fit is L2, with $\chi^2=19.09$, while the L7 fit
  gives $\chi^2=319.82$. The SED of the source and the two fits are
  plotted in Figure \ref{L2orL7}. Optical spectroscopy of this source
  would clearly be useful, as noted also by \citet{geissler11}. The
  relatively high $\chi^2$ of the {\em photo-type} best fit would have
  marked it down as an object worthy of further study.

\item ULAS J2304+1301 was discovered by \citet{DayJones13} who
  classified it T0, from a near-infrared spectrum. The source is
  classified L3.5 by {\em photo-type} with a satisfactory
  $\chi^2=4.82$. In contrast, fitting the T0 template yields
  $\chi^2=216.24$, a very poor fit. The two fits are compared against
  the SED of the source in Figure \ref{L3.5orT0}. As can be seen, the
  source is substantially bluer than a T0 in both $i-z$ and
  $W1-W2$. The source has $i-z=2.18\pm0.13$ (Vega), and is visible as
  the T0 outlier in Fig. \ref{polyiz}. This compares to the expected
  colour $i-z=2.15$ for an L3.5 dwarf, and $i-z=3.36$ for a T0 dwarf
  (Table \ref{tab:template}). We have checked the SDSS images which
  appear normal.

It is possible the source is an unresolved binary, and spectroscopy in
the $iz$ region would be revealing. Nevertheless given the fact that a
single L3.5 provides a satisfactory fit to the photometry, this could
be an example of an unresolved binary that cannot be identified from
colours alone, as prefigured in $\S\ref{subsec:binaries}$. It provides a warning that if a
spectrum over a limited wavelength range provides a substantially
different classification to the {\em photo-type} classification, the
source may be an unresolved binary.

\end{itemize}

\subsubsection{SpeX follow up}
\label{accuracy:spex}
A cleaner estimate of the accuracy of {\em photo-type} may be obtained
from uniform high-quality spectra over a wide wavelength range, using
the same instrumental set-up, to ensure uniform accurate spectral
classifications. For this purpose we selected a sample of objects from
the catalogue of Paper II for follow-up observations. We limited the
sample to sources with $\chi^2<15$, for the fits to the 8-band
photometry, to avoid outliers (objects with large $\chi^2$ are
considered in detail in Paper II). Other than that we selected objects
at random from a range of spectral types.

\begin{figure} 
\includegraphics[width=9.0cm]{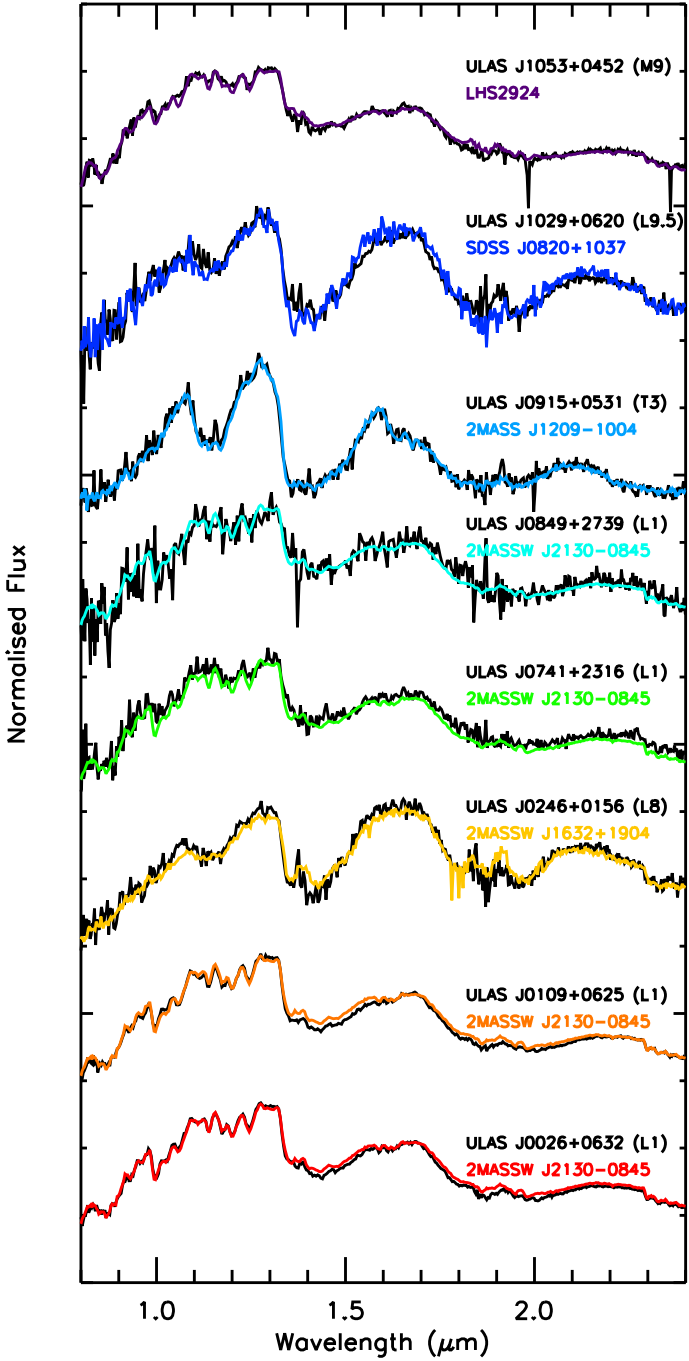} 
\caption{SpeX spectra of eight sources used to estimate the accuracy
  of {\em photo-type}. The sources are classified as L1 (ULAS
  J0026+0632), L1 (ULAS J0109+0625), L8 (ULAS J0246+0156), L1 (ULAS J0741+2316), L1 (ULAS
  J0849+2739), T3 (ULAS J0915+0531), L9.5 (ULAS J1029+0620), and M9 (ULAS
  J1053+0452), by comparison against the spectroscopic standards listed
in Table \ref{spec_standards}. The standard closest in spectral type is
also shown for each source.}
\label{good_spectra} 
\end{figure}

We obtained spectra of the 8 sources listed in Table
\ref{good_sources} with the SpeX instrument on the NASA Infrared
Telescope Facility between March and November 2013. The data were
reduced using the SpeXtool package version 3.4 \citep{cushing04}. The
spectra are plotted in Figure \ref{good_spectra}.

The spectra were classified by one of us (JKF), by visual comparison
against the SpeX Prism Spectral Library maintained by AJB, and without
knowledge of the {\em photo-type} classifications. The standards are
listed in Table \ref{spec_standards}. The resulting spectroscopic
classifications are listed in Table \ref{good_sources}, where they are compared to the
{\em photo-type} classifications. All objects are confirmed as ultra
cool dwarfs. Taking
the spectroscopic classification to be the correct classification, the
accuracy of {\em photo-type} estimated from this small sample is only
0.4 sub-types {\em r.m.s.}

In all, including peculiar objects (Paper II), so far we have observed
20 objects from our list of L and T dwarfs, and all are ultra-cool
dwarfs. This indicates, at least, that the contamination of the L and
T sample of Paper II is not large. Nevertheless a larger spectroscopic
sample would be required to quantify this accurately.

\subsubsection{Simulated data}
\label{accuracy:simulated}
\begin{figure} 
\includegraphics[width=9.0cm]{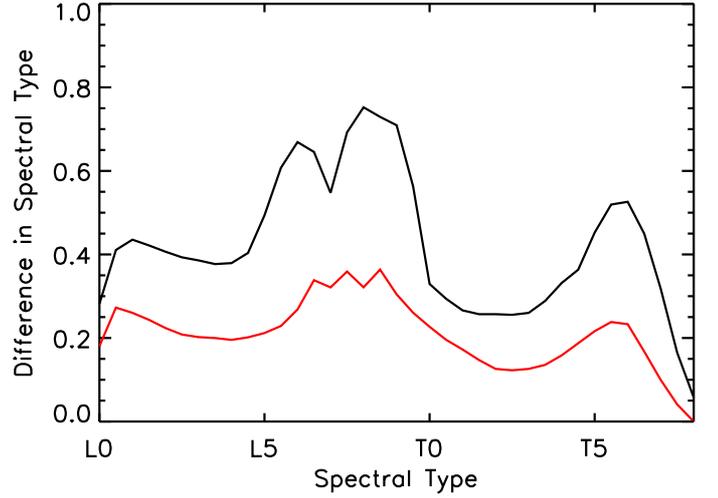} 
\caption{Estimated accuracy {\em r.m.s.} of the {\em photo-type}
  classification for sources of $J=16.0$ (red curve) and $J=17.5$
  (black curve) based on classification of synthetic data.}
\label{scattermcmc} 
\end{figure}

A third estimate of the accuracy of {\em photo-type} was obtained by
Monte Carlo methods. For a particular $J$ magnitude and for each
spectral type, we created synthetic data, accounting as appropriate
for the photometric errors and the intrinsic scatter in the colours of
the population (by adding an error of 0.05 magnitudes in each band in
quadrature to the photometric error, \S\ref{subsec:templates}). Then we classified every
synthetic object, and measured the dispersion in the classification
about the input spectral type. In Figure \ref{scattermcmc} we show the
measured dispersion for $J=16.0$, and at the
sample limit $J=17.5$. This analysis indicates that the classification method is accurate to better than one spectral type, for all spectral types, even at the sample limit, $J=17.5$. The plot shows that the method performs least
well, around spectral type L6, as expected (see discussion in \S\ref{subsec:templates}),
because several of the colour relations are relatively flat around
this spectral type. 

We can use the same apparatus to quantify the usefulness of the different photometric bands, by simply removing one or more bands, and observing the effect on the classification accuracy. We found that the $i$ band contributes usefully to the classification of L dwarfs, but that for our dataset the $i$ band makes a negligible contribution to the classification of T dwarfs, because the photometric errors are so large. The WISE data are useful in classifying all spectral types. The improvement in the classification accuracy depends on spectral type and brightness, but on average including the WISE data reduces the uncertainty in the spectral type by $\sim30\%$.

Finally we looked at the effect on the classification accuracy of disregarding the uncertainties of the polynomial fits. We found that including or excluding this error term has a negligible effect on the accuracy of the classification for our SDSS/UKIDSS/WISE dataset. For most bands and spectral types this is because the fit error is smaller than the intrinsic colour scatter, as shown in Fig. \ref{covar_err}. Even the large fit errors for the $i-z$ and $K-W1$ colours for late T dwarfs do not influence the classification. The reason for this is that these colours are in any case unimportant in classifying late T dwarfs, where the $W1-W2$ colour makes the main contribution.
 
The three estimates of accuracy are in reasonable agreement, and indicate
that brighter than $J=17.5$, for single normal L and T dwarfs, the
accuracy of {\em photo-type} classifications is at the level of one
spectral sub-type r.m.s.

\section{{\em photo-type} cookbook}
\label{sec:cook}

In this section, as a reference, we provide a brief summary of how to
use {\em photo-type} to classify a photometric source, with complete or partial
photometry from the $izYJHKW1W2$ photometric dataset. All photometry
is assumed to be on the Vega system.

It is not necessary to have photometry in all the bands. Neither is it
necessary to compute any colours explicitly. The steps involved are to
compute $\chi^2$ for each template, and then select the template with
the smallest $\chi^2$ as the classification.

For the object, add 0.05 magnitude intrinsic scatter in quadrature to
the photometric error in each band. Then, for a particular template,
setting $J=0$, and using the colours in Table \ref{tab:template} (or
using the polynomial relations), assemble the template magnitudes for
the bands in which photometry is available. Next, compute the magnitude
offset that provides the minimum $\chi^2$ match of the template to the
object, from equation \ref{weighted_mean}, and calculate
$\chi^2$ for that template using equation \ref{chi2_eqn}. Repeat the
procedure for all templates to obtain the classification, as the
template with the minimum $\chi^2$. The $\chi^2$ of the best fit
provides an indication of whether the object is peculiar or not.

The accuracy of the classification will depend on the brightness of
the source and the number of photometric bands. The accuracy may be
estimated by the Monte Carlo method of the previous section. Starting
with the measured photometry for the object, create a large number of
synthetic objects by adding errors in each band drawn from a Gaussian
of the appropriate dispersion (i.e. random + intrinsic scatter). Then
classify each of these synthetic objects as if real objects, and
record the scatter. For this purpose it makes sense to classify to the
nearest half spectral sub-type in order to measure the dispersion more
accurately.

\section{Summary}
\label{sec:summary}

In this paper we have described a new method to identify
and accurately classify L and T dwarfs in multiwavelength 0.75-4.6 $\mu$m
photometric datasets. For typical L and T dwarfs the classification is
accurate to one spectral sub-type r.m.s. The sample of 1157 L and T
dwarfs, $13.0<J<17.5$, selected from an area of 3344 deg$^2$, is
provided and described in Paper II. The principal benefit of the {\em
  photo-type} method is the production of a sample of L and T dwarfs
across the entire range L0 to T8, with accurate spectral types, that
is an order of magnitude larger than previous homogeneous samples, and
therefore ideal for statistical studies of, for example, the
luminosity function, and for quantifying the dispersion in properties
for any particular sub-type. The {\em photo-type} method can also be
used to select unusual objects, including unresolved binaries, or rare
types, identified by large $\chi^2$. An important advantage of the
method is that it covers a broad wavelength range, 0.75 to 4.6 $\mu$m,
meaning that the method may identify peculiar objects that look normal
in spectra that cover only a small wavelength range, and so might
otherwise be overlooked. The strengths and weaknesses of {\em
  photo-type} for the study of unusual L and T dwarfs are discussed in
detail in Paper II.

\begin{acknowledgements}
We would like to thank the anonymous referee for several constructive comments that helped improve the paper. The authors are grateful to Subhanjoy Mohanty for discussions that
contributed to this work significantly. We also acknowledge a useful
discussion with Nicholas Lodieu who encouraged us to include objects
undetected in SDSS in the sample. NS would like to thank STFC for the
financial support supplied.\\
This research has benefited from the SpeX Prism Spectral Libraries,
maintained by Adam Burgasser at http://pono.ucsd.edu/~adam/browndwarfs/spexprism.\\
This publication makes use of data products from the Wide-field
Infrared Survey Explorer, which is a joint project of the University
of California, Los Angeles, and the Jet Propulsion
Laboratory/California Institute of Technology, and NEOWISE, which is a
project of the Jet Propulsion Laboratory/California Institute of
Technology. WISE and NEOWISE are funded by the National Aeronautics
and Space Administration.\\
This research has benefitted from the M, L,
T, and Y dwarf compendium housed at DwarfArchives.org. The UKIDSS
project is defined in \citet{UKIDSS}. UKIDSS uses the UKIRT Wide Field
Camera \citep{casali07}. The photometric system is described in
\citet{hewett06}, and the calibration is described in
\citet{hodgkin09}. The science archive is described in
\citet{hambly08}.\\
Funding for SDSS-III has been provided by the Alfred P. Sloan
Foundation, the Participating Institutions, the National Science
Foundation, and the U.S. Department of Energy Office of Science. The
SDSS-III web site is http://www.sdss3.org/.  SDSS-III is managed by
the Astrophysical Research Consortium for the Participating
Institutions of the SDSS-III Collaboration including the University of
Arizona, the Brazilian Participation Group, Brookhaven National
Laboratory, Carnegie Mellon University, University of Florida, the
French Participation Group, the German Participation Group, Harvard
University, the Instituto de Astrofisica de Canarias, the Michigan
State/Notre Dame/JINA Participation Group, Johns Hopkins University,
Lawrence Berkeley National Laboratory, Max Planck Institute for
Astrophysics, Max Planck Institute for Extraterrestrial Physics, New
Mexico State University, New York University, Ohio State University,
Pennsylvania State University, University of Portsmouth, Princeton
University, the Spanish Participation Group, University of Tokyo,
University of Utah, Vanderbilt University, University of Virginia,
University of Washington, and Yale University.

\end{acknowledgements}

\bibliographystyle{aa}
\bibliography{skrzypek_090714_article_final}


\begin{appendix}
\section{Bayesian template classification}
\label{section:method}

The classification scheme described in \S3\ is based on 
a simple best-fit $\chi^2$ statistic, but is motivated by a fully Bayesian
approach to photometric template-fitting.
The full Bayesian result is derived below, 
after which a series of approximations are made to 
obtain the $\chi^2$ classification scheme --
one benefit of at least starting with a 
Bayesian formalism is that all
assumptions and approximations must be made explicit.

For a target source with photometric measurements
$\{\est{m}_\band\}$ and uncertainties $\{\sigma_\band\}$
in each of $\nband$ passbands 
(\ie, $\band \in \{ 1 , 2, \ldots, \nband \}$),
the aim here is to evaluate the probability,
$\prob(t | \{\est{m}_\band\}, \{ \sigma_\band \}, \ntype )$,
that it is of type $t$,
given that there are $\ntype$ types of astronomical object
(indexed by $t \in \{1, 2, \ldots, \ntype\}$)
under consideration.
Under the assumption that the source is of one of these types, 
Bayes's theorem implies that 
\begin{equation}
\label{equation:bt}
\prob(t | \{\est{m}_\band\}, \{ \sigma_\band \}, \ntype)
  = \frac
    {
      \prob(t | \ntype ) \,
      \prob(\{\est{m}_\band \} | \{\sigma_\band \}, t)
    }
    {
      \sum_{t^\prime=1}^{\ntype} 
      \prob({t^\prime} | \ntype ) \,
      \prob(\{\est{m}_\band \} | \{\sigma_\band \}, {t^\prime}) 
    },
\end{equation}
where 
$\prob(t | \ntype )$ is the prior probability
of the $t$'th model 
(\ie, how common this type of astronomical object is)
and 
$\prob(\{\est{m}_\band \} | \{\sigma_\band \}, t)$
is the marginal likelihood\footnote{The marginal
likelihood is sometimes referred to as the model-averaged
likelihood or, especially in astronomy, as the (Bayesian) evidence.}
that the observed 
photometry would have been obtained for an object of type $t$.


Each type is assumed to be specified by a template
of model colours (\ie, band-to-band magnitude differences),
defined relative to some reference passband $B$.
The colour for template $t$ and band $b$ is
denoted $c_{b,t}$, with $c_{B,t} = 0$ by construction.
The specification of templates by colours
means that the source's (unknown) magnitude in the reference band,
$\offset_B$, must also be included in each model.
If this quantity is not of interest (as is the case here), 
$\offset_B$ can be
integrated out to give the marginal likelihood for the $t$'th type as 
\begin{equation}
\label{equation:evidence}
\prob(\data | \noise, t) 
  = \int_{-\infty}^\infty
    \prob(\offset_B | t) \,
    \prob(\data | \noise, \offset_B, t) \, \diff \offset_B,
\end{equation}
where $\prob(\offset_B | t)$ is the prior distribution 
of $\offset_B$ for objects of the $t$'th type
(and so approximately proportional to their observed number counts)
and 
$\prob(\data | \noise, \offset_B, t)$ is the likelihood
of obtaining the measured data given a value for $\offset_B$.

Under the assumptions
that the measurements in the $\nband$ bands are indepenent and
that the variance is additive and normally distributed
(in magnitude units\footnote{This is not a good approximation
for sources that are fainter than the detection limit in any of the 
relevant bands; in this case the likelihood should be calculated 
in flux units as described in, \eg, \cite{mortlock12}.}),
the likelihood is
\begin{equation}
\label{equation:lik_1}
\prob(\data | \noise, \offset_B, t)
  = 
  \prodband
   \frac{
   \exp\left[ - \frac{1}{2} 
     ( \est{m}_\band - \offset_B - c_{\band, t})^2 / \sigma_\band^2
    \right]
  }
  {
(2 \pi)^{1 / 2} \sigma_\band
  }
,
\end{equation}
where $\offset_B + c_{\band,t}$ is the predicted $\band$-band 
magnitude\footnote{While the template is specified in terms of colours, 
at no point are observed colours 
of the form $\hat{m}_b - \hat{m}_{b^\prime}$ ever calculated.
If observed colours were used then the resultant likelihood 
would have to incorporate the correlations induced by the fact that 
the same measured magnitude was used to calculate more than one colour.
The resultant likelihood could not be expressed
in terms of a simple $\chi^2$ statistic as is done here.}.
Equation~\ref{equation:lik_1} can be re-written as 
\begin{equation}
\label{equation:chisq}
\prob(\data | \noise, \offset_B, t)
  = 
   \frac{
\exp\left[ - \frac{1}{2} \chi^2(\data, \noise, \offset_B, t) \right]
    }
   {\prodband (2 \pi)^{1/2} \sigma_\band },
\end{equation}
where
\begin{equation}
\label{equation:chi2}
\chi^2(\data, \noise, \offset_B, t) 
  = \sumband 
   \left( 
     \frac{\est{m}_\band - \offset_B - c_{\band, t} }{\sigma_\band} 
   \right)^2 
\end{equation}
is the standard $\chi^2$ mis-match statistic.
For the purposes of evaluating the integral in equation~\ref{equation:evidence}
it is useful to further rearrange equation~\ref{equation:lik_1} 
into the form 
\begin{equation}
\label{equation:lik_2}
\prob(\data | \noise, \offset_B, t)
\end{equation}
\[
\mbox{}
=
\frac{
  \exp\left[
  -\frac{1}{2}
\chi^2(\data, \noise, \est{\offset}_{B,t}, t) 
   \right]
}
{
\prodband (2 \pi)^{1 / 2} \sigma_\band
}
  \exp
    \left\{ -\frac{1}{2}
    \left[
      \frac{ \offset_B - \est{\offset}_{B,t} }
  {\left(\sumband 1/\sigma_\band^2\right)^{-1/2}} \right]^2
   \right\},
\]
where
\begin{equation}
\label{equation:offsetest}
\est{\offset}_{B,t}
  = \frac{\sumband ({\est{m}_\band - c_{\band,t}}) / \sigma_\band^2}
  {\sumband 1 / \sigma_\band^2}
\end{equation}
is the 
natural inverse-variance weighted estimate of $\offset_B$
for this combination of source photometry and template,
and 
$\chi^2(\data, \noise, \est{\offset}_{B,t}, t)$ is similarly the 
minimum value of $\chi^2$.

Inserting the above expression for the likelihood into
equation~\ref{equation:evidence} allows the marginal
likelihood to be written as 
\[
\prob(\data | \noise, t)
=
\frac{
  \exp\left[
  -\frac{1}{2}
\chi^2(\data, \noise, \est{\offset}_{B,t}, t) 
   \right]
}
{
\prodband (2 \pi)^{1 / 2} \sigma_\band
}
\]
\begin{equation}
\label{equation:evidence2}
\times 
\int_{-\infty}^\infty
  \prob(\offset_B | t) \, 
  \exp
    \left\{ -\frac{1}{2}
    \left[
      \frac{ \offset_B - \est{\offset}_{B,t} }
  {\left(\sumband 1/\sigma_\band^2\right)^{-1/2}} \right]^2
   \right\} \, \diff \offset_B,
\end{equation}
illustrating that the 
goodness of fit 
and the number counts of this type 
play quite strongly separated roles in this problem.

The classification statistic defined 
in \S3\ comes from adopting
a uniform $\offset_B$ prior distribution 
of the form 
\begin{equation}
\prob(\offset_B | t) 
  = \step(\offset_{B} - \offset_{B,\minimum})
  \, \step(\offset_{B,\maximum} - \offset_{B}) 
  \frac{1}{\offset_{B,\maximum} - \offset_{B,\minimum}},
\end{equation}
where $\step(x)$ is the Heaviside step function
and 
$\offset_{B,\minimum}$ and $\offset_{B,\maximum}$
are taken to be the same for all types 
(hence the lack of a $t$ subscript).
Provided that 
$\offset_{\minimum} \ll \hat{\offset}_{B,t}$
and 
$\offset_{\maximum} \gg \hat{\offset}_{B,t}$,
the integral in 
equation~\ref{equation:evidence2} can be approximated analytically to give
the marginal likelihood as 
\[
\prob(\data | \noise, t) 
\]
\begin{equation}
\mbox{}
  = 
    \frac{
    \left(\sumband 1 / \sigma_\band^2\right)^{-1/2}
    }
    {(2 \pi)^{\nband / 2 - 1} \left( \prodband \sigma_\band \right)}
   \frac{
  \exp\left[
  -\frac{1}{2}
  \chi^2(\data, \noise, \est{\offset}_t, t)
  \right]}
  {\offset_{B,\maximum} - \offset_{B,\minimum}},
\end{equation}
where 
$\est{\offset}_{B,t}$ is given in equation~\ref{equation:offsetest}
and 
$\chi^2(\data, \noise, \offset_B, t)$ is given in 
equation~\ref{equation:chisq}.
Inserting this expression into equation~\ref{equation:bt},
the posterior probability that the source is of type $t$ becomes
\[
\prob(t | \data, \noise, \ntype )
\]
\begin{equation}
\mbox{}
  = \frac
    {
      \prob(t | \ntype )  \,
\exp\left[
  -\frac{1}{2}
  \chi^2(\data, \noise, \est{\offset}_{B,t}, t)
  \right] 
    }
    {
      \sum_{t^\prime=1}^{\ntype} 
      \prob({t^\prime} | \ntype )  \,
\exp\left[
  -\frac{1}{2} 
  \chi^2(\data, \noise, \est{\offset}_{B,t^\prime}, {t^\prime})
  \right]
    }.
    \label{equation:final}
\end{equation}

This probabilistic template matching scheme can be made absolute by
classifying a source as being of the type with the maximum
posterior probability, 
which in turn corresponds to the maximum value of the numerator of \ref{equation:final},
$\prob(t | \ntype )  \,
\exp[ - \chi^2(\data, \noise, \est{\offset}_{B,t}, t) / 2 ]$.
If the relative numbers of the different source types are comparable
(or if the templates have very distinct colours, relative to the 
photometric noise)
then the differences in the priors can be neglected, 
in which case a source would be classified as being of the type t
which yields the minimum value of 
$\chi^2(\data, \noise, \est{\offset}_{B,t}, t)$.
This is the approach taken in \S2.
%

\end{appendix}


\end{document}